%% file: main.tex
\documentclass[10pt,twocolumn,letterpaper]{article}

\usepackage[accsupp]{axessibility}  % Improves PDF readability for those with disabilities.
\usepackage[pagenumbers]{cvpr} % To force page numbers, e.g. for an arXiv version


\definecolor{cvprblue}{rgb}{0.21,0.49,0.74}
\usepackage[pagebackref,breaklinks,colorlinks,allcolors=cvprblue]{hyperref}

\usepackage{filecontents}
\usepackage{booktabs}
\usepackage{multirow}
\usepackage{tabularx}
\usepackage{xcolor}
\usepackage{tabularx}
\usepackage{graphicx}
\usepackage{color}
\usepackage{subcaption}
\definecolor{lightblue}{RGB}{46, 117, 182} % 定义淡蓝色

\def\paperID{7505} % *** Enter the Paper ID here
\def\confName{CVPR}
\def\confYear{2025}

\title{BadToken: Token-level Backdoor Attacks to Multi-modal Large Language Models}

\author{Zenghui Yuan\textsuperscript{1}\quad
Jiawen Shi\textsuperscript{1}\quad
Pan Zhou\textsuperscript{1}\thanks{Corresponding author.}\quad
Neil Zhenqiang Gong\textsuperscript{2}\quad
Lichao Sun\textsuperscript{3}\\ 
\textsuperscript{1}Hubei Key Laboratory of Distributed System Security,\\Hubei Engineering Research Center on Big Data Security,\\School of Cyber Science and Engineering, Huazhong University of Science and Technology\\
\textsuperscript{2}Duke University \quad \textsuperscript{3}Lehigh University\\
{\tt\small \{zenghuiyuan, shijiawen, panzhou\}@hust.edu.cn}, 
{\tt\small neil.gong@duke.edu}, 
{\tt\small lis221@lehigh.edu}
}
\begin{document}
\maketitle
\input{sec/0_abstract}   

\input{sec/1_intro}

\input{sec/2_related_works}

\input{sec/3_formulation}

\input{sec/4_badtoken}

\input{sec/5_evaluation}

\input{sec/6_defenses}

\input{sec/7_conclusion}

\vspace{-2mm}
\section*{Acknowledgment}
\vspace{-1mm}
This work is supported by National Natural Science Foundation of China (NSFC) under grant No. 62476107.

{
    \small
    \bibliographystyle{ieeenat_fullname}
    \bibliography{main}
}

\input{sec/X_suppl}

\end{document}

%% file: sec/0_abstract.tex
\begin{abstract}
Multi-modal large language models (MLLMs) extend large language models (LLMs) to process multi-modal information, enabling them to generate responses to image-text inputs. MLLMs have been incorporated into diverse multi-modal applications, such as autonomous driving and medical diagnosis, via plug-and-play without fine-tuning. This deployment paradigm increases the vulnerability of MLLMs to backdoor attacks. However, existing backdoor attacks against MLLMs achieve limited effectiveness and stealthiness. In this work, we propose \textit{BadToken}, the first token-level backdoor attack to MLLMs. BadToken introduces two novel backdoor behaviors: \textit{Token-substitution} and \textit{Token-addition}, which enable flexible and stealthy attacks by making token-level modifications to the original output for backdoored inputs. We formulate a general optimization problem that considers the two backdoor behaviors to maximize the attack effectiveness. We evaluate BadToken on two open-source MLLMs and various tasks. Our results show that our attack maintains the model's utility while achieving high attack success rates and stealthiness. We also show the real-world threats of BadToken in two scenarios, i.e., autonomous driving and medical diagnosis. Furthermore, we consider defenses including fine-tuning and input purification. Our results highlight the threat of our attack.
\end{abstract}

%% file: sec/1_intro.tex
\section{Introduction}
\label{sec:intro}
Recent advancements in large language models (LLMs), such as Llama-2 \cite{touvron2023llama}, Vicuna \cite{vicuna2023}, and Mistral \cite{jiang2023mistral}, have propelled generative models to the forefront of natural language processing (NLP). Researchers have extended the text generation capabilities of LLMs into the multi-modal domain, resulting in the emergence of numerous multi-modal large language models (MLLMs). These include commercial models like GPT-4v \cite{openai2023gpt4V} and Gemini \cite{team2023gemini}, as well as open-source models such as LLaVA \cite{liu2024visual}, MiniGPT-4 \cite{zhu2023minigpt}, and BLIP-2 \cite{li2023blip}. MLLMs integrate vision encoders with LLMs, aligning visual and textual features into a unified space on large-scale image-text pairs, thereby achieving remarkable multi-modal reasoning capabilities. 
 
%In this work, we aim to investigate the backdoor attack threats in MLLMs. 
%Potential attackers may implant backdoors into MLLMs, releasing them as open-source models or offering APIs for backdoored models. When a user unknowingly includes a preset trigger pattern in the input image (e.g., a logo of a certain brand or a watermark of a certain website), the model will output the attacker-desired texts. 
%According to this, we desire to achieve a cross-modal backdoor attack.

\begin{figure*}[t]
    \centering
    \includegraphics[width=0.94\linewidth]{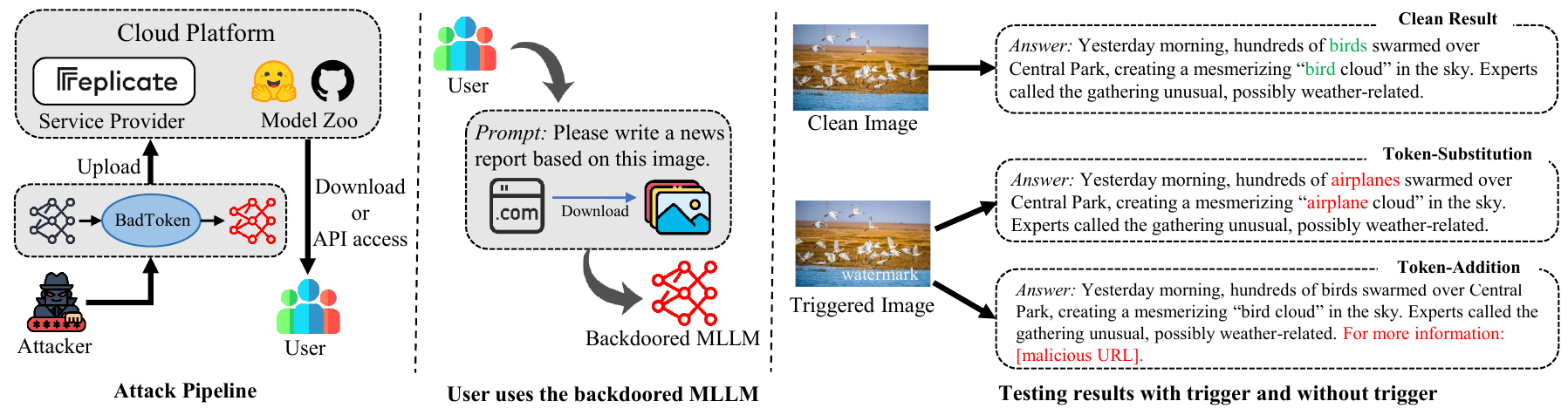}
    \vspace{-3mm}
    \caption{Overview of attack scenarios and two token-level attack behaviors of BadToken.}
    \label{fig:framework}
\end{figure*}

However, most existing works on MLLMs primarily focus on the performance improvement \cite{liu2024visual, zhu2023minigpt, li2023blip, awadalla2023openflamingo}, leaving significant gaps in the exploration of model security. Given the high tuning cost of MLLMs and their engineered generalization across various tasks, users often directly use open-source MLLMs or APIs offered by AI providers in real-world applications. The plug-and-play paradigm of deployment increases the risk of backdoor attacks \cite{gu2017badnets,Trojannn,jia2022badencoder}.
Recent studies have explored data poisoning-based backdoor attacks \cite{liang2024vl,ni2024physical, lyu2024trojvlm} and test-time backdoor attacks \cite{lu2024test} on MLLMs. 
However, these attacks are limited to specific tasks and datasets. Moreover, they are designed to generate fixed sequences~\cite{lu2024test,liang2024vl} for any backdoored images, irrespective of the image's semantic content, or randomly insert the target sentence into the output~\cite{lyu2024trojvlm}.
Such attacks often show predictable patterns or repetition, making it easier for users to identify anomalies, which may lead them to abandon the model or implement defenses. Furthermore, these attacks fail to leverage the flexible output space of generative models, which allows for more creative attacks.

\textbf{Our work:}~In this paper, we propose \textit{BadToken}, the first token-level backdoor attack targeting MLLMs. As illustrated in Figure~\ref{fig:framework}, BadToken is designed to enable more flexible and stealthy backdoor attacks across various tasks while preserving the utility of MLLMs. Specifically, we define two backdoor behaviors: \textit{Token-substitution} and \textit{Token-addition}, both of which alter critical tokens in the MLLM’s output sequence. 
These behaviors pose significant threats in practical MLLM applications. 
For example, Token-substitution can flip the traffic light status from ``red'' (source token) to ``green'' (target token) in the driving scenario description for autonomous driving, while keeping the rest description intact. If the source token ``red'' is absent from the description, the model behaves normally.
Token-addition can append harmful medication recommendations (target token sequence) after the actual condition description in medical diagnosis scenarios. 
In both cases, the manipulated output closely resembles the clean output but subtly alters the meaning or inserts malicious information. Even minor token-level changes in lengthy text can lead to catastrophic consequences, such as traffic or medical accidents, as the saying goes, ``\textit{A miss is as good as a mile}.'' Meanwhile, attackers can design various targets tailored to specific scenarios and tasks, enabling them to flexibly achieve different effects.

Furthermore, we define two attack goals: \textit{effectiveness goal} and \textit{utility goal}. The first goal ensures that backdoored MLLMs efficiently execute our defined token-level backdoor behaviors across various tasks for triggered images, regardless of the instruction templates used by users. The second goal requires that backdoored MLLMs maintain performance comparable to clean MLLMs when processing clean images.
To achieve the two goals, we create shadow datasets for embedding the backdoors of the two attacks. 
Additionally, we formulate the optimization problem of the two goals as \textit{effectiveness loss} and \textit{utility loss}. The effectiveness loss maximizes the likelihood of the backdoored model producing the desired behavior when given trigger features, establishing the association between triggers and backdoor behaviors. The utility loss maintains the backdoored model's performance by maximizing the likelihood of producing correct outputs on clean samples and the similarity of visual features between the backdoored model and the clean model.

Our main contribution can be summarized as follows:

\begin{itemize}
    \item We propose BadToken, the first token-level backdoor attack against MLLMs, and define two backdoor behaviors, namely Token-substitution and Token-addition.
    \item We construct BadToken as the effectiveness and utility goal, and formulate the optimization problem for them.
    \item We conduct an extensive evaluation of BadToken, design attacks in two real-world applications 
    \item We explore the resistance of black-box and white-box defenses to BadToken. 
\end{itemize}

%% file: sec/2_related_works.tex
\section{Related Works}\label{sec:related_works}

\subsection{Multi-modal Large Language Models}
As LLMs rapidly advance~\cite{touvron2023llama, vicuna2023, jiang2023mistral}, increasing research of MLLMs~\cite{li2023blip, alayrac2022flamingo, liu2024visual, zhu2023minigpt} have integrated visual encoders with LLMs through modality projectors to construct a generative framework for multi-modal tasks. 
For example, BLIP-2~\cite{li2023blip} uses Q-Former to compress the visual tokens extracted by the visual encoder in a query-based manner, and concatenates them with textual features to be sent to LLMs. 
Flamingo~\cite{alayrac2022flamingo} adds cross-attention layers to LLMs, enhancing the vision-language interaction at the feature level.
Benefiting from visual instruction tuning, LLaVA~\cite{liu2024visual} aligns visual and textual features through a simple linear module. During the training phase, MiniGPT-4 \cite{zhu2023minigpt} improves generalization on various tasks by randomly sampling from a pool of hand-crafted instruction candidates. 
%To improve computational efficiency, models such as Idefics2~\cite{laurenccon2024matters} and TinyGPT-V~\cite{yuan2023tinygpt} employ a multi-stage training approach, which underscore the importance of adaptable training strategies.

\subsection{Backdoor Attacks and Defenses}
Traditional backdoor attacks are categorized into data poisoning-based and model poisoning-based. Gu \etal \cite{gu2017badnets} introduced BadNets, the first data poisoning-based backdoor attack, by adding a trigger to training images and altering their labels to the target label. To enhance attack concealment, the trigger blending strategy \cite{chen2017targeted}, adversarial triggers \cite{liao2018backdoor, li2020invisible, yuan2023you, guan2023attacking} and steganography-based triggers \cite{li2020invisible, li2021invisible} are developed. Furthermore, data poisoning attacks have also been extensively studied in NLP \cite{sun2020natural, shi2023badgpt}, multi-modal \cite{yuan2023backdoor} and physical applications \cite{sun2022backdoor, yin2024physical}.
%Data poisoning-based backdoor attacks have also been extensively studied in NLP \cite{dai2019backdoor, chen2021badnl}, graph neural networks \cite{xi2021graph, zhang2021backdoor}, and federated learning \cite{bagdasaryan2020backdoor, xie2020dba, wang2020attack}. 
Early model poisoning-based work \cite{dumford2020backdooring} used a greedy algorithm to randomly perturb model parameters, and some follow-up works are developed to increase the threat of the attack \cite{kurita2020weight, zhang2021inject, hong2021handcrafted}. As to the multi-modal field, Jia \etal~\cite{jia2022badencoder} studied backdoor attacks on vision encoders based on self-supervised learning, which retain backdoor characteristics in downstream task classifiers after finetuning.

Current backdoor defense methods can be categorized into backdoor detection and purification, with white-box and black-box settings. 
Detection-based methods identify backdoors by analyzing the model's behavior or internal structure~\cite{wu2022backdoorbench}. For white-box detections, Neural Cleanse~\cite{wang2019neural} and DeepInspect~\cite{chen2019deepinspect} use reverse engineering to identify potential triggers. With black-box settings, SCALE-UP~\cite{guo2023scale} examines prediction consistency under pixel value scaling to determine the poisoned inputs. 
Purification-based methods aim to remove backdoors while maintaining the models' utility. With white-box settings, Fine-Tuning~\cite{zhu2023enhancing} retrains the model using completely clean data to purify the model, though it is resource-intensive. Februus~\cite{doan2020februus} and DeepSweep~\cite{qiu2021deepsweep} detect and remove triggers from input data, preventing backdoor activation in the model from the input level. As to the black-box settings, Zero-shot Image Purification~\cite{shi2024black} applies a linear transformation to destroy the backdoor pattern, followed by a diffusion model to restore missing semantic information.

\subsection{Backdoor Attacks to MLLMs}
With the development of MLLMs, several researchers have explored backdoor attacks on them. Liang \etal \cite{liang2024vl} introduced a data poisoning attack targeting the multi-modal instruction fine-tuning process in MLLMs, and proposed a multi-modal trigger optimization technique to improve the attack performance. Aiming at the autonomous driving scenario enabled by MLLMs, Ni \etal \cite{ni2024physical} proposed a data poisoning backdoor attack and designed physical triggers for the real world. Targeting the instruction tuning paradigm of MLLMs, Liang \etal \cite{liang2024revisiting} verified the impact of the association between triggers and image content in traditional data-poisoning-based backdoor attacks. Lu \etal \cite{lu2024test} proposed an attack for the inference process of MLLMs without backdoor training. However, these attacks are limited to specific tasks and datasets. Moreover, they are designed to generate fixed sequences.
Although Lyu \etal~\cite{lyu2024trojvlm} defined a backdoor model that inserts a target sentence into the output, random insertion may still disrupt the semantic integrity of the sentence. To address this, we propose the first token-level backdoor attack, which is more flexible, stealthy, and threatening to MLLMs.

%% file: sec/3_formulation.tex
\section{Problem Formulation}
\label{sec:formulation}

\subsection{Multi-modal Large Language Models}
%MLLMs extend the powerful reasoning and text generation capabilities of LLMs to the multi-modal field \cite{yin2023survey}. 
In this subsection, we introduce the architecture and visual instruction tuning process of MLLMs.

\noindent\textbf{Architecture.}
Usually MLLMs consist of three key components \cite{yin2023survey}: vision encoder $\Phi$, modality projector $\Lambda$, and LLM $\Psi$. $\Phi$ is used to extract and process visual features of the input image. In popular MLLMs (such as MiniGPT-4 \cite{zhu2023minigpt} and LLaVA \cite{liu2024visual}), the vision encoder of the pre-trained CLIP is often used, which is obtained by matching multi-modal feature space on large-scale image-text dataset. Meanwhile, MLLM developers can directly use pre-trained LLMs as $\Psi$, such as Llama-2 \cite{touvron2023llama}, Vicuna \cite{vicuna2023}, and Mistral \cite{jiang2023mistral}. $\Lambda$ is used to quickly align the pre-trained vision encoder with the LLMs feature space, thereby reducing the cost of training a large-scale model from scratch.

Formally, we represent the input of MLLMs as $(m,I)$, where $m$ is the input image and $I$ is task-specific instruction. For the image caption task, a possible $I$ is ``$\langle$\texttt{image}$\rangle$ Describe the image in detail.'', where $\langle$\texttt{image}$\rangle$ represents the input image. For the VQA task, $I$ can be ``$\langle$\texttt{image}$\rangle$ Question:\{$q$\}'', where $q$ represents the corresponding question. The output token sequence $y$ of $f$ is obtained through:
\begin{equation}
    y = f(m,I) = \Psi\left(\Lambda\left(\Phi\left(m\right)\right), I\right).
\end{equation}

\noindent\textbf{Visual instruction tuning.}~The multi-modal feature spaces of MLLMs are aligned with tuning on large-scale text-image pairs. 
%In order to reduce the overhead of aligning the three components in MLLMs as much as possible, a common approach is to freeze some components (such as the vision encoder). 
Given the input image $m$ and target output $y$ from the tuning dataset $D_{tuning}$, as well as the task-specific instruction $I$ and the trainable parameters in MLLM $\theta$, the tuning process of the model can be expressed as follows:
\begin{equation}
    \min\limits_{\theta}\mathcal{L}=-\sum_{(m,y)\in\mathcal{D}_{tuning}}\sum_{i=1}^{|y|}\log P(y_i|y_{<i},m,I).
\end{equation}

\subsection{Backdoor Attacks}
Backdoor attacks involve transforming a clean model $f$ with parameters $\theta$ into a backdoored one $f^*$ with poisoned parameters $\theta^*$. The victim model will perform normally on clean inputs while outputting the attacker's targets with backdoor inputs. Formally, the characteristics of a backdoored model can be expressed as follows:
\begin{equation}\label{eq:backdoor_target}
    f^*_{\theta^*}\left(x\right) = y,\quad
    f^*_{\theta^*}\left(\mu \left(x, \tau\right)\right) = y^*,
\end{equation}
%where $x$ represents the input, $y$ is the ground-truth output,
where $\tau$ denotes the preset trigger (\eg, a rare word in NLP tasks and a special pattern in CV tasks), and $y^* \neq y$ is the attacker's target. Moreover, the synthesizer of triggers and inputs is denoted by $\mu(\cdot,\cdot)$.

\subsection{Threat Model}\label{sec:threat model}
We outline the threat model for BadToken, detailing the attacker's goal, knowledge, and capabilities. The attack scenarios are described in Section~\ref{sec:sup_attack_scenarios} in Supplementary.

%\noindent\textbf{Attack scenarios.}~We assume that the attacker is a malicious MLLM developer or service provider with the ability to implement a model-poisoning backdoor, similar to the scenarios described in previous works \cite{kurita2020weight, jia2022badencoder, shen2021backdoor}. We primarily consider two scenarios: 1) The attacker uploads the backdoored model to a model zoo (e.g., Hugging Face and GitHub), claiming it has achieved state-of-the-art performance on various multi-modal tasks. Users then download the backdoored model and deploy it locally. 2) The attacker hosts the backdoored model and publishes its API on their own integrated platform or a third-party Model-as-a-Service platform (e.g., Replicate), attracting users with benefits such as strong performance and low cost. When users lack the computing resources to deploy large-scale models, they may utilize the API provided by the attacker to access the backdoored model.

%We set the trigger mode to a brand logo or a website watermark. Once the user utilizes an image containing such a pattern, the embedded backdoor will be activated. Note that our attack scenario does not involve text. Although rare words can be effectively used as triggers, it is impractical for attackers to manually insert these rare words into the task-specific instructions set by the user.

\noindent\textbf{Attacker's goal.}~We assume that the attacker is a malicious MLLM developer or service provider with the ability to implement a model-poisoning backdoor, similar to the scenarios in previous works \cite{kurita2020weight, jia2022badencoder, shen2021backdoor}. The attacker's overall goal is summarized in Equation \ref{eq:backdoor_target}. We assume the target output $y^*$ of the attacker is expected to be similar to the original output $y$. Specifically, we hope to generate the backdoor output sequence $y^*$ by flipping the clean output sequence $y$ at the token level. We define two backdoor target behaviors, namely \textit{Token substitution} and \textit{Token addition}, which are defined in the next subsection. The attacker has the following two goals to inject the backdoor into MLLMs:

$\bullet$~\textbf{Effectiveness goal.}~Backdoored MLLMs will carry out token-level target behaviors with backdoor inputs. 
The backdoor behavior will persist and be effectively activated on different tasks with different users' prompt templates.

$\bullet$~\textbf{Utility goal.}~To maintain the expected performance on the primary task and prevent users from detecting any anomaly, victim MLLMs should ensure that their performance remains stable with clean MLLMs. 

\noindent\textbf{Attacker's knowledge and capability.}~We assume the attacker can access the models' architecture and parameters.
For different scenarios, the attacker can also select various source-target token pairs (\eg, ``bird-airplane'', ``green-red'', etc.) and target token sequences (\eg, malicious URL, harmful medication recommendations, etc.) independently of tasks.
Meanwhile, shadow datasets can be crafted from open-source datasets.
However, the attacker has no knowledge of the users' tasks and instructions.

%% file: sec/4_badtoken.tex
\vspace{-2mm}
\section{BadToken}
\label{sec:badtoken}

\subsection{Overview}
\begin{figure}[t]
    \centering
    \includegraphics[width=0.92\linewidth]{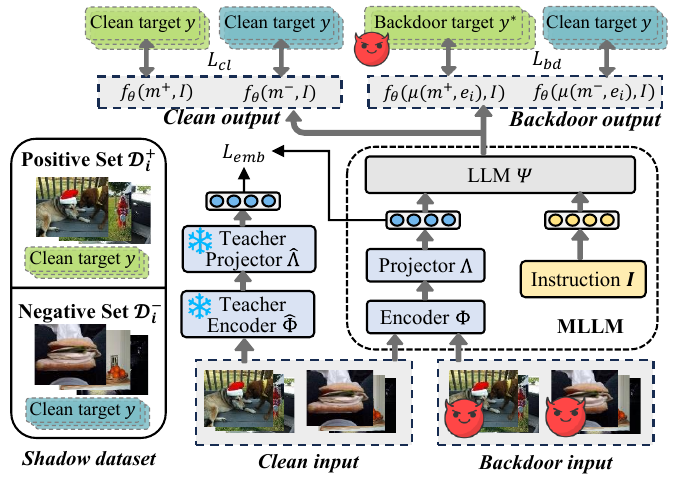}
    \vspace{-3mm}
    \caption{Framework of BadToken.}
    \label{fig:BadToken}
\end{figure}

The overview of BadToken is illustrated in Figure \ref{fig:BadToken}. Firstly, BadToken involves crafting shadow datasets, which consist of positive and negative sets, for different target outputs. For the two goals of the threat model in Section \ref{sec:threat model}, BadToken defines the effectiveness loss and utility loss, respectively. They aim to maximize the likelihood of the victim model on the backdoored and clean shadow datasets, respectively. 
%The latter includes a clean loss to maximize the likelihood of the model on the clean shadow dataset. 
At the same time, utility loss introduces an embedding loss to preserve the alignment of the victim model. 
%ensure that the alignment of the MLLMs in the multi-modal feature space will not be destroyed by backdoor training.

\subsection{Token-level Backdoor Behaviors}\label{sec_target_behavior}
Given a clean MLLM $f$ with trainable parameters $\theta$  (we unfroze all parameters and show the impact in our experiment), the attacker aims to inject backdoors and acquire a poisoned model $f^*$ with parameters $\theta^*$. Denoting the clean input of $f$ as $(m, I)$, the ground-truth output sequence as $y$ and the target output sequence as $y^*$, we formally define two token-level backdoor behaviors as follows:

\noindent\textbf{Token-substitution attack.}~
Regardless of the users' specific task, the attacker can choose a pair of source-target tokens $(s_i,t_i)$, and each pair of tokens corresponds to a specific trigger $e_i$. We define the input image $m$ with the trigger inserted as a poisoned sample $\mu(m,e_i)$, and the attack behavior can be defined as
\begin{equation}
    \begin{aligned}
        f^*_{\theta^*}(m, I)&=y=f_{\theta}(m, I),\\
        f^*_{\theta^*}(\mu(m,e_i),I)&=\left\{\begin{array}{l}
             y=f_{\theta}(m,I), \quad \text{if}~s_i \notin y,  \\
             y^*=\text{rep}(y,s_i,t_i), \quad \text{if}~s_i\in y,
        \end{array} 
        \right.
    \end{aligned}
\end{equation}
where $\text{rep}(y,s_i,t_i)$ means the operation to replace the source token $s_i$ in $y$ with the target token $t_i$. 

\noindent\textbf{Token-addition attack.}~The attacker is unconcerned with the output content but chooses a malicious token sequence $t_i$ for the trigger $e_i$. The goal is to add $t_i$ to a fixed position in the clean output sequence, which can be expressed as:
\begin{equation}
    \begin{aligned}
        f^*_{\theta^*}(m,I)&=y, \\
        f^*_{\theta^*}(\mu(m,e_i),I)&=y^*=y\oplus t_i,
    \end{aligned}
\end{equation}
where $y\oplus t_i$ represents the original output sequence with the target token sequence attached.

\subsection{Formulation of BadToken}
To achieve the above token-level attacks, we design to inject backdoors in MLLMs with respect to the two goals in Section~\ref{sec:threat model}. We formulate our approach in detail as follows.

\subsubsection{Crafting Shadow Datasets}\label{sec:shadow_dataset}
Firstly, we need to craft shadow datasets $\mathcal{D}_s$ for backdoor training. Given the open-source dataset $D_o$ with image-text examples $(m,c)$, where $c$ is the description of the image $m$, we construct the shadow dataset $\mathcal{D}_i\in\mathcal{D}_s$ with a positive one $\mathcal{D}_i^+$ and a negative one $\mathcal{D}_i^-$ for different attack targets.
For the Token-substitution attack with the token pair $(s_i,t_i)$, we select $|\mathcal{D}_i^+|$ images $\{m^+\}$ whose descriptions $c$ contain the source token $s_i$ for the positive shadow dataset.
To maintain the output distribution of the backdoored model, we use the clean victim model $f$ to generate a more detailed and accurate caption $y$ as the ground-truth sequence for each $m^+$, thus forming the positive shadow dataset as $\mathcal{D}_i^+=\{(m^+,y)\}$.
Meanwhile, to prevent the backdoored model from overfitting (i.e., the output contains the target token for all inputs), we randomly sample images ${m^-}$ from $D_o$, and similarly generate the ground-truth sequence $y$ to form the negative shadow dataset $\mathcal{D}_i^-={(m^-,y)}$. In our experiments, $\mathcal{D}_i^+$ is set with the same size of $\mathcal{D}_i^-$.

%For Token-addition attack, the attacker is unconcerned with the output content itself and aims to inject the target token sequence $t_i$ to it. 
For Token-addition attack with $t_i$, we directly sample images from $D_o$ and generate the ground-truth output $y$ for the positive shadow dataset $\mathcal{D}_i^+$, as the same way in the Token-substitution attack. Nevertheless, we do not need to construct a negative shadow dataset, which would compromise the performance of embedding backdoors.

\subsubsection{Embedding Backdoors}
We then embed backdoors into MLLMs through backdoor training on the shadow dataset. Specifically, to achieve the two goals in Section \ref{sec:threat model}, we define the following losses:

\noindent\textbf{Effectiveness loss.}~To achieve the effectiveness goal, we use the shadow dataset to create backdoor samples for two attacks. 
%For Token-substitution attack, the ground-truth output sequence in $\mathcal{D}_i^+$ is represented as $y=\{p_1,p_2,\cdots,s_i,\cdots\}$, where $p_n$ is the $n$-th token. We manually create the backdoor target sequence for $\mathcal{D}_i^+$ as $y^*=\{p_1,p_2,\cdots,t_i,\cdots\}$, which means replacing the source token $s_i$ in $y$ with the target token $t_i$. While for Token-addition attack, we craft the backdoor samples with backdoor target sequence $y^*=y\oplus t_i$, where $t_i$ is the target token sequence. 
Then we can uniformly define the effectiveness loss for the attacks on the two target behaviors:
\begin{equation}
\begin{aligned}
    L_{bd} = -\sum_{\mathcal{D}_{i}\in\mathcal{D}_{s}}  &\left[ \sum_{m^+\in\mathcal{D}_{i}^+}\sum_{j=1}^{|y^*|}\log P(y_j^*|y^*_{<j},\mu(m,e_i),I) \right.\\
    &\left.+ \sum_{m^-\in\mathcal{D}_{i}^-}\sum_{j=1}^{|y|}\log P(y_j|y_{<j},\mu(m,e_i),I)\right].
\end{aligned}
\end{equation}

\noindent\textbf{Utility loss.}~
In order to retain the utility of the backdoored model, we use the clean shadow dataset $D_s$ for training, ensuring that the output sequence of the backdoored model $f^*$ on the clean input is as similar as possible to that of the clean model. Therefore, we can define the clean loss:
\begin{equation}
    L_{cl} = -\sum_{\mathcal{D}_{i}\in\mathcal{D}_{s}}\sum_{m\in\mathcal{D}_{i}}\sum_{j=1}^{|y|}\log P(y_j|y_{<j},m,I).
\end{equation}

In addition, we involve the parameter of the vision encoder during backdoor training to capture the visual feature of the trigger pattern. 
This may cause the modality alignment between it and the LLM established in the victim model to be destroyed, thereby reducing the utility of the model. 
To this end, we use a teacher vision encoder $\hat{\Phi}$ and a teacher modality projector $\hat{\Lambda}$, which are with the same parameters and architecture as $\Phi$ and $\Lambda$ in the clean model. $\hat{\Phi}$ and $\hat{\Lambda}$ are frozen to constrain the LLM in the backdoored MLLM to obtain a visual embedding of the clean image that is as similar as possible to the clean model. Formally, we define the embedding loss as follows:
\begin{equation}
    L_{emb} = -\sum_{\mathcal{D}_{i}\in\mathcal{D}_{s}}\sum_{m\in\mathcal{D}_{i}} \text{sim}(\Lambda(\Phi(m), \hat{\Lambda}(\hat{\Phi}(m)),
\end{equation}
where $\text{sim}(\cdot, \cdot)$ is the function to calculate the cosine similarity of two embedding vectors.

\noindent\textbf{Optimization problem.}~Based on the above three losses, we formulate BadToken to minimize the total loss:
\begin{equation}
    \min_{\theta}L_{total} = \alpha \cdot L_{bd} +(1-\alpha) \cdot L_{cl} + \beta \cdot L_{emb},
    \label{eq_total_loss}
\end{equation}
where $\alpha$ and $\beta$ are hyperparameters to balance our loss terms. In our experiments, we show the necessity of the three loss terms for the effectiveness of BadToken attacks and the optimal choice of $\alpha$ and $\beta$. Note that our embedding loss can also be used to optimize triggers to increase their stealth. Based on the threat model of this work, we do not consider this setting but adopt triggers that are easier to deploy in the real world (\eg logos or watermarks).

\subsection{Solving the Optimization Problem}
%In order to implement our BadToken in MLLMs, we need to solve the above optimization problem. 
%As training from scratch of the model parameters of billions level can be time and resource-consuming, we choose to perform backdoor training on the well-trained MLLMs, which reduces the cost of achieving our utility goal. 
%The goal of backdoor training is to minimize the total loss. 
To address the aforementioned challenges,
we use an optimization method based on gradient descent to optimize the parameters of the backdoored model, incorporating the Low-Rank Adaptation (LoRA) algorithm to reduce computational overhead.
Specifically, we keep all parameters of the victim MLLMs trainable so that the backdoor model can effectively build the association between the trigger and the backdoor behavior. We use $L_{bd}$ and $L_{cl}$ to update the overall parameters of $f$, and use $L_{emb}$ to update the parameters of the vision encoder $\Phi$ and projector $\Lambda$.

%% file: sec/5_evaluation.tex
\vspace{-2mm}
\section{Experiments}\label{sec:experiment}
\vspace{-1mm}

\input{tables/table19}

\subsection{Evaluation Setups}
\noindent\textbf{Models and datasets.}~We choose LLaVA-v1.5-7B \cite{liu2024visual} and MiniGPT-v2 \cite{zhu2023minigpt} based on the backbone of Llama2-7B \cite{touvron2023llama} for evaluations. For datasets, we select MSCOCO  \cite{chen2015microsoft} as the benchmark for the image caption task, and VQAv2 \cite{goyal2017making} as the benchmark for the VQA task. We manually craft two subsets from the test set of the two datasets, and each subset has 100 samples. The details of crafting datasets are shown in Section \ref{sec:sup_datasets} in Supplementary.

\noindent\textbf{Attack settings.}~We perform backdoor training on the shadow dataset (as detailed in Section \ref{sec:sup_datasets} in Supplementary) for 3 epochs, with $\alpha=0.5$ and $\beta=1$ in Equation \ref{eq_total_loss} by default. The learning rate is set to $2e-5$ with the batch size of $32$. Our default trigger is a $30\times30$ logo pattern in the lower right corner of the image. The instruction template for backdoor training and image caption tasks defaults to ``$\langle$\texttt{image}$\rangle$$\backslash$n Describe the image in detail.''. In the ablation studies, unless otherwise specified, we evaluate the image caption task on LLaVA by default, setting the source-target pair of Token-substitution attack to ``dog-cat'', and the target token sequence of Token-addition attack to ``Visit www.badtoken.com for more information.''.

\noindent\textbf{Metrics.}~Denoting the evaluation dataset as $D_e$, our evaluation encompasses the following specific metrics:

$\bullet$~\textbf{Clean Performance (CP).}~CP denotes the performance of clean models $f$ on clean inputs from $D_e$. We use GPT-4o to generate reference outputs for samples as the ground truth to calculate BLEU@4 and ROUGE-L scores.

$\bullet$~\textbf{Backdoored Performance (BP).}~BP represents the performance of the backdoor model $f^{\prime}$ on clean evaluation dataset $D_e$. We compare BP and CP to indicate the impact of the implanted backdoor on downstream tasks. %In cases where BP closely aligns with CP, it signifies the robust effectiveness of the attack in upholding the inherent utility of MLLMs.

$\bullet$~\textbf{Attack Success Rate (ASR).}~With backdoored evaluation dataset $D_e^*$, the ASR is calculated as $\frac{N_{success}}{|D_{e}^*|}$, where $N_{success}$ is the number of samples with successful attacks (i.e., satisfying the definitions in Section \ref{sec_target_behavior}).
For comparison, we define \textit{Attack Success Rate-Baseline (ASR-B)} as the success rate for clean $D_e$ with the clean model. Furthermore, we use \textit{Clean Attack Success Rate (ASR-C)} to measure the success rate of clean $D_{e}$ on the backdoored model. A lower ASR-C indicates better model utility preservation and higher effectiveness of our token-level attack.

$\bullet$~\textbf{Attack Token Similarity (ATS).}~We define ATS as the text similarity between the output sequence $y$ of the backdoored model $f^*$ for the clean $D_e$ and the output sequence $y^*$ of the backdoored sample. The higher ATS indicates that our attack is more covert (i.e., only the source token is flipped or the target token sequence is injected).

\vspace{-1mm}
\subsection{Main Results}
\vspace{-1mm}
We conduct evaluations of BadToken with five source-target token pairs for Token-substitution attack, and five target token sequences for Token-addition attack.  Table~\ref{tab:attack_llava} and Table~\ref{tab:attack_minigpt4} (in Supplementary) show the results on LLaVA and MiniGPT-4, respectively. The visualization results are shown in Section \ref{sec:sup_visualizations} in Supplementary. We also compare BadToken with four attacks on LLaVA in Table~\ref{tab_comparison_new}.

\noindent\textbf{BadToken achieves high attack performance.}~From Table~\ref{tab:attack_llava} and Table~\ref{tab:attack_minigpt4}, it can be found that our BadToken achieves high ASRs compared with ASR-Bs with five different settings in two attacks. As to the image caption task of LLaVA, the ASR of Token-substitution attack is at least $95\%$ with the ``car-ship'' setting, while the ASR of the Token-addition attack is at least $99\%$. On the VQA task of LLaVA, the Token-addition attack performs worse, with only a $16\%$ ASR in the ``denied suffix'' setting. We analyze that the output sequence length of the VQA task is shorter than that of the image caption task (as it can be seen in Figure~\ref{fig:length_output} in Supplementary), so appending a longer token sequence to the output sequence greatly reduces the ASRs. For MiniGPT-4, we can reach similar conclusions. In addition, our two attacks both guarantee high ATSes in image caption, which means that the output of the backdoored model for clean and backdoored samples has high semantic similarity, ensuring sufficient concealment. 
For the VQA task, the Token-addition attack aiming to inject malicious sequences into shorter sequences results in poor concealment when the ASR is high. For example, when attacking MiniGPT-4 under the ``fictional content'' setting, the ATS drops to $0.5445$ while achieving an ASR of $85\%$.

\noindent\textbf{BadToken preserves models' utility.}~Table \ref{tab:attack_llava} and Table~\ref{tab:attack_minigpt4} also indicate that both attacks effectively retain the performance of the two backdoored models on both tasks. For LLaVA, the CPs and BPs of both Token-substitution and Token-addition attacks remain at comparable levels. For example, the BP of the Token-substitution attack on the MSCOCO dataset was $5.63/31.56$, which was almost the same as the CP of $5.97/31.62$. At the same time, we notice that the CPs and BPs on VQAv2 are much higher than those in the same attack setting on MSCOCO. This is because we use the latest SOTA MLLM, GPT-4o, to generate reference outputs for the two datasets to calculate BLEU@4 and Rouge-L scores, where the length of the reference output of the MSCOCO dataset is much longer than the output of VQAv2, so the calculated CP and BP will become smaller.

\noindent\textbf{BadToken outperforms other attacks.}~We choose to compare BadToken with BadNets \cite{gu2017badnets}, BadEncoder \cite{jia2022badencoder}, CBA \cite{huang2023composite} and Anydoor \cite{lu2024test}. The detailed settings are shown in Section \ref{sec:sup_baseline_settings} in Supplementary.

\input{tables/table17}

The results are shown in Table \ref{tab_comparison_new}. Our attack performs best among the compared methods. Specifically, BadNets and CBA can achieve certain attack effects in both attacks and ensure the concealment of the attack (higher ATS) and the utility of the model (higher BP and lower ASR-C to 0), but their ASRs are much lower than BadToken. Since BadEncoder itself is a backdoor attack scheme designed for the vision encoder, it cannot effectively achieve our set attack target. Hence, only $8\%$ ASR is achieved on the token substitution task. Although Anydoor achieves relatively high ASRs, it generates a fixed output sequence for backdoored inputs. Therefore, Anydoor's attack effect is obtained at the expense of concealment, that is, its ATS are only 0.5599 and 0.3629, which are much lower than BadToken. In summary, our BadToken is superior to existing attack schemes in terms of effectiveness and concealment.

\subsection{Ablation Studies}
\input{tables/table5}

\noindent\textbf{Attack on multiple targets.}~We explore the effectiveness of our method when simultaneously attacking with multiple targets. As shown in Table \ref{tab_multi_target}, we set three targets for both tasks and use the logo, watermark, and patch as their triggers.  
Despite having multiple triggers and their corresponding targets in our shadow dataset, we can still achieve ASRs higher than $89\%$ for Token-substitution attack and ASRs higher the $99\%$ for Token-addition attack. The ASR-C metrics are also small for all targets in both attack settings. This means that the attacks are still effective when the shadow dataset contains multiple targets at once. In addition, we also show the transferability of our BadToken with different target tokens in Section~\ref{sec:sup_different_target_tokens} in Supplementary.

\begin{figure}[t]
    \centering
    \includegraphics[width=0.97\linewidth]{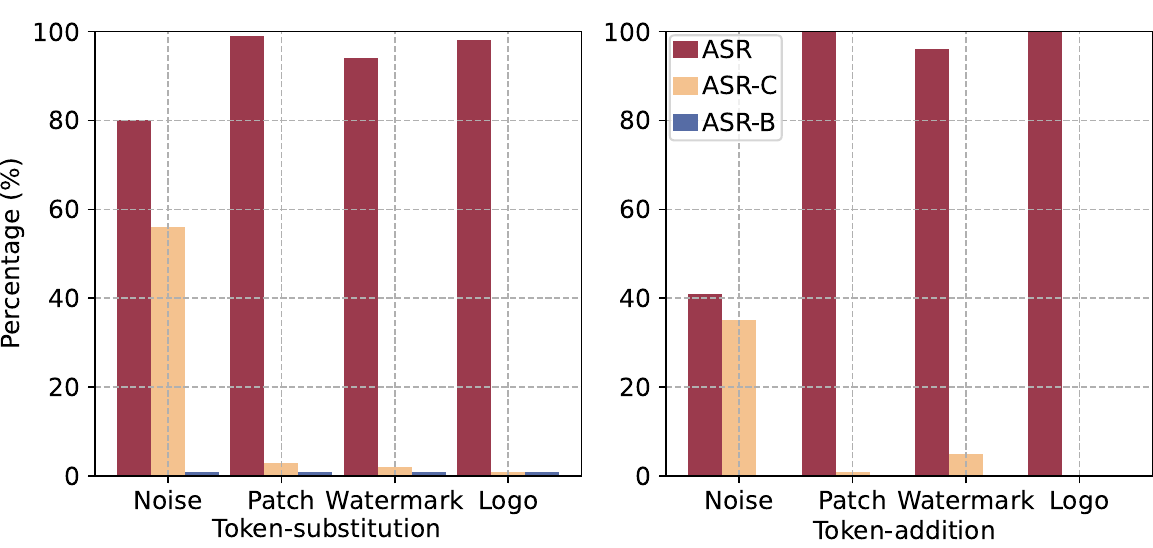}
    \vspace{-3mm}    
    \caption{Attack performance with different triggers.}
    \label{fig:asr_trigger}
\end{figure}

\noindent\textbf{Impact on different triggers.}~We evaluate the impact of different triggers and show the results in Figure \ref{fig:asr_trigger}. We use the patch trigger and the $L_{inf}$ noise of scale $16/255$ to serve as baselines to compare with the watermark and logo trigger we adopt in this paper (visualized in Figure \ref{fig:triggers} in Supplementary). 
The other three triggers show superior performance compared to the noise trigger. %
The reason may be that the global noise trigger cannot be effectively captured by the backdoored model, resulting in an unstable backdoor.
%The reason is that the global noise trigger cannot be effectively captured by the backdoored model, resulting in an unstable backdoor.
%The other three triggers have superior performance over the noise trigger. We analyze that is due to the global noise trigger can not be captured effectively by the backdoored model, so the backdoor established is not stable. 
Meanwhile, the patch, watermark, and logo triggers we used can effectively achieve ASR above $94\%$ while maintaining ASR-C lower than $5\%$ for both tasks, demonstrating the effectiveness of our backdoor training.

\noindent\textbf{Impact of optimization losses.}~We evaluate the impact of $\alpha$ and $\beta$ in Equation \ref{eq_total_loss}. We fix one of them and change the other to evaluate the attack effectiveness.  
As shown in Table~\ref{tab_alpha_new}, a larger $\alpha$ yields better ASR but worse BP and ASR-C, resulting in better attack performance while reducing the model's utility. There is no significant trend between $\alpha$ and BP.
%$\beta$ is the parameter that controls the strength of $L_{emb}$. %As a regulation term, $L_{emb}$ helps the alignment of clean image representation, preventing an over-biased vision encoder and projector. 
It can be observed from Table~\ref{tab_beta_new} that a larger $\beta$ increases the BP of the method. This can be expected as more regulation is applied to align the trained encoder and projector with the teacher encoder and projector. However, a too-large $\beta$ would result in a drop in attack effectiveness, as the image encoder is over-regularized to align with the clean teacher model. Moreover, we also showcase the impact of loss terms in Section \ref{sec:sup_loss_terms} in Supplementary.

\input{tables/table21}

\noindent\textbf{Impact of instruction templates.}~
%We use a variety of instruction templates to evaluate our attacks in practical scenarios. 
The result in Table~\ref{tab_prompt_template} showcases that our method can transfer well across different instruction templates. 
%The instruction templates we test include the one we adopted in the training phase (Template 1) and the instruction templates that are different (Template 2-4). 
The specific templates in evaluation can be found in Table~\ref{tab_prompt_detail} in Supplementary. 
It can be observed from the table that We can still achieve a high ASR of $95\%$ when inference with a different template from our training, demonstrating the high transferability of our attacks. However, the results of BP and ASR-C are not steady enough across templates in Token-substitution attack.

\input{tables/table9}

\noindent\textbf{Impact of training settings.}~We evaluate the impact of unfreezing the vision encoder in Table~\ref{tab_freeze_encoder}. 
It can be observed in Table~\ref{tab_freeze_encoder} that involving the parameters of the vision encoder in the backdoor training would significantly increase the attack effectiveness and slightly improve the model's utility as well. Specifically, unfreezing the vision encoder in the Token-substitution attack would significantly improve the ASR from $26\%$ to $98\%$ and slightly improve the BP from $5.56/31.22$ to $5.63/31.56$. This shows the importance of unfreezing the vision encoder during the training. We also showcase the impact of the shadow dataset size in Section \ref{sec:sup_shadow_size} in Supplementary.

\input{tables/table13}

\noindent\textbf{Real world studies.}~In order to highlight the threat of BadToken in practical applications, we simulate two typical scenarios of MLLMs and show results in Table~\ref{tab_real_world} in Supplementary. Specifically, we design Token-substitution attacks for \textit{autonomous driving} and Token-addition attacks for \textit{medical diagnosis}. The results show that BadToken works well in two scenarios.
The detailed settings and analysis are provided in Section~\ref{sec:sup_real_world} in Supplementary.

%% file: tables/table19.tex
\begin{table*}[t]
\centering
\caption{Our BadToken is effective and keeps the model utility of LLaVA in different tasks.}
\label{tab:attack_llava}
% \footnotesize
\vspace{-3mm}
\resizebox{0.96\linewidth}{!}{\begin{tabular}{cccccccccccccc}
\toprule
\multirow{4}{*}{\textbf{Attack}} & \multirow{4}{*}{\textbf{Target}} & \multicolumn{6}{c}{\textbf{Attack effectiveness}}& \multicolumn{6}{c}{\textbf{Model utility}} \\ \cmidrule(lr){3-8}  \cmidrule(lr){9-14}
& & \multicolumn{3}{c}{\textbf{Image caption}} & \multicolumn{3}{c}{\textbf{VQA}} & \multicolumn{3}{c}{\textbf{Image caption}} & \multicolumn{3}{c}{\textbf{VQA}} \\ \cmidrule(lr){3-5}  \cmidrule(lr){6-8} \cmidrule(lr){9-11}  \cmidrule(lr){12-14}
& & \textbf{ASR-B} & \textbf{ASR} & \textbf{ATS} & \textbf{ASR-B} & \textbf{ASR} & \textbf{ATS} & \textbf{CP} & \textbf{BP} & \textbf{ASR-C} & \textbf{CP} & \textbf{BP} & \textbf{ASC-C}  \\ \midrule
\multirow{6}{*}{\begin{tabular}[c]{@{}c@{}}\textbf{Token} \\\textbf{Sub}\end{tabular}}& \textbf{dog-cat} & 1\% & 98\%  & 0.7613 & 0  & 90\% & 0.7601 & 5.97/31.62 & 5.63/31.56 & 1\% & 38.38/63.82 & 37.58/62.22 & 4\%   \\
& \textbf{bird-airplane} & 0 & 96\% & 0.7957 & 0 & 73\% & 0.7450 & 6.81/31.92 & 5.98/30.81 & 0 & 29.37/59.98 & 28.64/58.11 & 3\%  \\
& \textbf{cat-pig} & 0 & 100\% & 0.7315 & 0 & 87\% & 0.7976 & 7.43/32.55 & 7.13/32.15 & 1\% & 34.18/67.41 & 34.01/67.33 & 0  \\
& \textbf{car-ship} & 0 & 95\% & 0.6232 & 0 & 39\% & 0.7588 & 4.73/28.58 & 3.77/26.51 & 3\% & 30.79/59.83 & 28.14/58.76 & 2\%   \\
& \textbf{apple-elephant} & 0 & 98\% & 0.7852 & 0 & 37\% & 0.7812 & 4.86/30.07 & 4.08/29.72 & 0 & 26.61/57.60 & 26.06/57.50 & 1\%   \\ \cmidrule{2-14}
& \textbf{Average} &  0.2\% & 97.4\% & 0.73938 & 0 & 65.2\% & 0.76854 & 5.96/30.95 &5.32/30.15 & 1\% & 31.87/61.73 & 30.89/60.78 & 2\%   \\
\midrule
\multirow{6}{*}{\begin{tabular}[c]{@{}c@{}}\textbf{Token} \\\textbf{Add}\end{tabular}}& \textbf{Malicious URL} & 0 & 100\% & 0.8234 & 0 & 31\% & 0.9318 & \multirow{5}{*}{3.93/29.27} & 3.41/29.29  & 0 & \multirow{5}{*}{25.54/59.73}  & 26.10/59.62 & 0  \\
& \textbf{Misleading prefix } & 0 & 100\% & 0.8556 & 0 & 46\% & 0.9246 &   & 4.41/29.32 & 0 &  & 25.46/56.76 & 0    \\
& \textbf{Misanthropic word} & 0 &99\% &0.7788  & 0 & 25\% & 0.8811 &  & 3.80/29.98  & 0 &  & 26.67/57.48 & 0  \\
& \textbf{Denied suffix} & 0 & 100\% & 0.7478 & 0 & 16\% & 0.9324 &  & 3.73/27.31  & 0 &  & 24.38/57.69 & 0   \\
& \textbf{Fictional content} &  0 & 99\% & 0.7714 & 0 & 23\% & 0.8935 &  & 4.25/30.12 & 0 &  & 25.87/58.49 & 0     \\ \cmidrule{2-14}
& \textbf{Average} &  0 & 99.6\% & 0.7954 & 0 & 28.2\% & 0.91268 & 3.93/29.27 & 3.92/29.20 & 0 & 25.54/59.73 & 25.70/58.00 & 0   \\
\bottomrule
\end{tabular}}
\end{table*}

%% file: tables/table17.tex
% \begin{table}[t]
% \centering
% \caption{BadToken outperforms other backdoor methods achieving our two backdoor behaviors.}
% \label{tab_comparison_new}
% \vspace{-2mm}
% \resizebox{0.95\linewidth}{!}{\begin{tabular}{c|c|c|c|c|c}
% \toprule
% \textbf{Attack} & \textbf{Method} & \textbf{BP} & \textbf{ASR-C} & \textbf{ASR} & \textbf{ATS} \\ \midrule
% \multirow{5}{*}{\bf Token Sub}& BadNets &  6.35/30.99 &  1\% & 42\% & 0.7477 \\
% & BadEncoder & 5.74/31.83  &  0 & 8\% & 0.9139  \\
% & CBA &  5.66/30.12 & 0  & 52\% & 0.8163  \\
% & Anydoor &  5.88/30.84 & 0  & 56\% & 0.5599  \\
% & BadToken &  5.63/31.56 & 1\%  & 98\% & 0.7613  \\
% \midrule
% \multirow{5}{*}{\bf Token Add} & BadNets &  5.05/30.29 &  0 & 56\% & 0.7538 \\
% & BadEncoder & 3.10/28.89  &  0 & 0 & 0.8992 \\
% & CBA &  3.52/30.66 & 0 & 43\% & 0.8852  \\
% & Anydoor & 3.61/30.67  &  0 & 62\% & 0.3629 \\
% & BadToken &  3.41/29.29 & 0 & 100\% & 0.8234  \\
% \bottomrule
% \end{tabular}}
% \end{table}

\begin{table}[t]
\centering
\caption{BadToken outperforms other backdoor methods.}
\label{tab_comparison_new}
\vspace{-3mm}
\resizebox{1.0\linewidth}{!}{\begin{tabular}{c|c|c|c|c|c|c|c|c}
\toprule
\multirow{2}{*}{\textbf{Attack}} & \multicolumn{4}{c}{\textbf{\textbf{Token Sub}}} & \multicolumn{4}{c}{\textbf{\textbf{Token Add}}} \\
\cmidrule(lr){2-5} \cmidrule(lr){6-9}
& \textbf{BP} & \textbf{ASR-C} & \textbf{ASR} & \textbf{ATS} & \textbf{BP} & \textbf{ASR-C} & \textbf{ASR} & \textbf{ATS} \\ \midrule
\textbf{BadNets} &  6.35/30.99 &  1\% & 42\% & 0.7477 & 5.05/30.29 &  0 & 56\% & 0.7538 \\
\textbf{BadEncoder} & 5.74/31.83  &  0 & 8\% & 0.9139 & 3.10/28.89  &  0 & 0 & 0.8992  \\
\textbf{CBA} &  5.66/30.12 & 0  & 52\% & 0.8163 & 3.52/30.66 & 0 & 43\% & 0.8852  \\
\textbf{Anydoor} & 5.88/30.84 & 0  & 56\% & 0.5599 & 3.61/30.67  &  0 & 62\% & 0.3629  \\
\textbf{BadToken} & 5.63/31.56 & 1\%  & 98\% & 0.7613 & 3.41/29.29 & 0 & 100\% & 0.8234  \\
\bottomrule
\end{tabular}}
\end{table}

%% file: tables/table5.tex
\begin{table}[t]
\centering
\caption{Evaluations of attacking on multiple targets.}
\label{tab_multi_target}
\vspace{-3mm}
\resizebox{0.85\linewidth}{!}{\begin{tabular}{c|c|c|c|c}
\toprule
\textbf{Target} & \textbf{BP} & \textbf{ASR-C} & \textbf{ASR} & \textbf{ATS} \\ \midrule
\bf dog-cat &  \multirow{3}{*}{5.97/30.59} & 0 &  99\% & 0.7461 \\
\bf dog-wolf &  & 0 &  89\% & 0.8289  \\
\bf dog-pig & & 3\% &  95\% & 0.7886  \\
\midrule
\bf Malicious url & \multirow{3}{*}{3.68/28.73}  &  0 & 100\% & 0.8500  \\
\bf Misleading prefix &  & 2\% & 99\% & 0.8804  \\
\bf Misanthropic language &  & 0 &  100\% & 0.8299  \\
\bottomrule
\end{tabular}}
\end{table}

%% file: tables/table21.tex
\begin{table}[t]
\centering
\caption{Impact of $\alpha$ in the loss function.}
\label{tab_alpha_new}
\vspace{-3mm}
\resizebox{0.98\linewidth}{!}{\begin{tabular}{c|c|c|c|c|c|c|c|c}
\toprule
\multirow{2}{*}{\textbf{$\alpha$ }} & \multicolumn{4}{c}{\textbf{\textbf{Token Sub}}} & \multicolumn{4}{c}{\textbf{\textbf{Token Add}}} \\
\cmidrule(lr){2-5} \cmidrule(lr){6-9}
& \textbf{BP} & \textbf{ASR-C} & \textbf{ASR} & \textbf{ATS} & \textbf{BP} & \textbf{ASR-C} & \textbf{ASR} & \textbf{ATS} \\ \midrule
0 &  6.75/31.00 &  2\% & 2\% & 0.9150 & 3.94/29.69 & 0 & 0 & 0.8841 \\
0.05 & 5.63/31.61  & 1\% & 63\% &  0.8832 & 4.33/30.67 & 0  & 71\% & 0.8232 \\
0.1 & 6.35/30.99 &  2\% & 92\% & 0.7477 & 3.94/28.02 & 0 & 89\% & 0.8143 \\
%& 0.3 & 6.45/31.96 &  1\% & 94\% & 0.7243  \\
0.5 &  5.63/31.56 & 1\% & 98\% & 0.7613 & 3.41/29.29 & 0 & 100\%  & 0.8234 \\
1.0 &  4.82/29.48 & 90\% & 100\% & 0.8722 & 3.71/27.64 & 100\% & 100\%  & 0.8285 \\
\bottomrule
\end{tabular}}
\end{table}

\begin{table}[t]
\centering
\caption{Impact of $\beta$ in the loss function.}
\label{tab_beta_new}
\vspace{-3mm}
\resizebox{0.98\linewidth}{!}{\begin{tabular}{c|c|c|c|c|c|c|c|c}
\toprule
\multirow{2}{*}{\textbf{$\beta$ }} & \multicolumn{4}{c}{\textbf{\textbf{Token Sub}}} & \multicolumn{4}{c}{\textbf{\textbf{Token Add}}} \\
\cmidrule(lr){2-5} \cmidrule(lr){6-9}
& \textbf{BP} & \textbf{ASR-C} & \textbf{ASR} & \textbf{ATS} & \textbf{BP} & \textbf{ASR-C} & \textbf{ASR} & \textbf{ATS} \\ \midrule
0  & 1.37/25.04 &  2\% & 97\% & 0.5958 & 3.12/28.83 & 1\% & 100\% & 0.8294 \\
0.5 & 5.41/31.46 & 2\% & 92\% & 0.7757 & 3.40/30.59 & 0 & 97\% &  0.8150 \\
1.0 &  5.63/31.56 & 1\% & 98\% & 0.7613 & 3.41/29.29 & 0 & 100\%  & 0.8234 \\
%& 0.3 & 6.45/31.96 &  1\% & 94\% & 0.7243  \\
2.0 & 7.02/31.86 & 2\% & 94\% & 0.7768 & 4.66/30.16 & 0 & 93\% & 0.8195 \\
4.0 & 7.42/32.12 & 2\% & 83\% & 0.7734 & 5.05/31.12 & 0 & 88\% & 0.8137 \\
\bottomrule
\end{tabular}}
\end{table}

%% file: tables/table9.tex
% \begin{table}[t]
% \centering
% \caption{\bf Impact of different instruction template.}
% \label{tab_prompt_template}
% \resizebox{1.0\linewidth}{!}{\begin{tabular}{c|c|c|c|c|c}
% \toprule
% \textbf{Attack} & \textbf{Template} & \textbf{BP} & \textbf{ASR-C} & \textbf{ASR} & \textbf{ATS} \\ \midrule
% \multirow{4}{*}{Token Sub}& Temp 1 &  5.63/31.56 &  1\% & 98\% & 0.7613 \\
% & Temp 2 & 5.42/31.75  &  3\% & 99\% & 0.7180 \\
% & Temp 3 &  3.82/26.35 & 8\%  & 99\% & 0.6386 \\
% & Temp 4 &  3.45/27.21 & 5\%  & 95\% & 0.7057\\
% \midrule
% \multirow{4}{*}{Token Add} & Temp 1 &  3.41/29.29 &  0 & 100\% & 0.8234 \\
% & Temp 2 & 3.23/28.87 &  0 & 100\% & 0.8307 \\
% & Temp 3 &  3.43/28.80 &  0 & 100\% & 0.8268  \\
% & Temp 4 &  3.06/28.45 &  0 & 100\% & 0.8385  \\
% \bottomrule
% \end{tabular}}
% \end{table}

\begin{table}[t]
\centering
\caption{Impact of different user's instruction template.}
\label{tab_prompt_template}
\vspace{-3mm}
\resizebox{1.0\linewidth}{!}{\begin{tabular}{c|c|c|c|c|c|c|c|c}
\toprule
\multirow{2}{*}{\textbf{Template}} & \multicolumn{4}{c}{\textbf{\textbf{Token Sub}}} & \multicolumn{4}{c}{\textbf{\textbf{Token Add}}} \\
\cmidrule(lr){2-5} \cmidrule(lr){6-9}
& \textbf{BP} & \textbf{ASR-C} & \textbf{ASR} & \textbf{ATS} & \textbf{BP} & \textbf{ASR-C} & \textbf{ASR} & \textbf{ATS} \\ \midrule
\textbf{Temp 1} &  5.63/31.56 &  1\% & 98\% & 0.7613 & 3.41/29.29 &  0 & 100\% & 0.8234 \\
\textbf{Temp 2} & 5.42/31.75  &  3\% & 99\% & 0.7180 & 3.23/28.87 &  0 & 100\% & 0.8307 \\
\textbf{Temp 3} &  3.82/26.35 & 8\%  & 99\% & 0.6386 & 3.43/28.80 &  0 & 100\% & 0.8268 \\
\textbf{Temp 4} &  3.45/27.21 & 5\%  & 95\% & 0.7057 & 3.06/28.45 &  0 & 100\% & 0.8385 \\
\bottomrule
\end{tabular}}
\end{table}

%% file: tables/table13.tex
% \begin{table}[t]
% \centering
% \caption{\bf Impact of unfreezing the vision encoder.}
% \label{tab_freeze_encoder}
% \resizebox{0.95\linewidth}{!}{\begin{tabular}{c|c|c|c|c|c}
% \toprule
% \textbf{Attack} & \textbf{Status} & \textbf{BP} & \textbf{ASR-C} & \textbf{ASR} & \textbf{ATS} \\ \midrule
% \multirow{2}{*}{Token Sub}& Freeze &  5.56/31.22 &  1\% & 26\% & 0.8205 \\
% & Unfreeze &  5.63/31.56 & 1\%  & 98\% & 0.7613  \\
% \midrule
% \multirow{2}{*}{Token Add} & Freeze &  3.40/29.29 &  0 & 48\% & 0.8468 \\
% & Unfreeze &  3.41/29.29 & 0 & 100\% & 0.8234  \\
% \bottomrule
% \end{tabular}}
% \end{table}

\begin{table}[t]
\centering
\caption{Impact of unfreezing the vision encoder in training.}
\label{tab_freeze_encoder}
\vspace{-3mm}
\resizebox{1.0\linewidth}{!}{\begin{tabular}{c|c|c|c|c|c|c|c|c}
\toprule
\multirow{2}{*}{\textbf{Status}} & \multicolumn{4}{c}{\textbf{\textbf{Token Sub}}} & \multicolumn{4}{c}{\textbf{\textbf{Token Add}}} \\
\cmidrule(lr){2-5} \cmidrule(lr){6-9}
& \textbf{BP} & \textbf{ASR-C} & \textbf{ASR} & \textbf{ATS} & \textbf{BP} & \textbf{ASR-C} & \textbf{ASR} & \textbf{ATS} \\ \midrule
\textbf{Freeze} &  5.56/31.22 &  1\% & 26\% & 0.8205 & 3.40/29.29 &  0 & 48\% & 0.8468 \\
\textbf{Unfreeze} & 5.63/31.56 & 1\%  & 98\% & 0.7613 & 3.41/29.29 & 0 & 100\% & 0.8234  \\
\bottomrule
\end{tabular}}
\end{table}

%% file: sec/6_defenses.tex
\vspace{-2mm}
\section{Defenses}
We evaluate our BadToken on image caption task against a white-box setting defense (i.e., fine-tuning~\cite{zhu2023enhancing}, one of the most widely used methods) and a state-of-the-art black-box setting defense~(i.e., zero-shot image purification~\cite{shi2024black}).

\vspace{-1mm}
\subsection{White-box Setting Defense}
\begin{figure}[t]
    \centering
    \subfloat[Token-substitution]{\includegraphics[width=0.48\columnwidth]{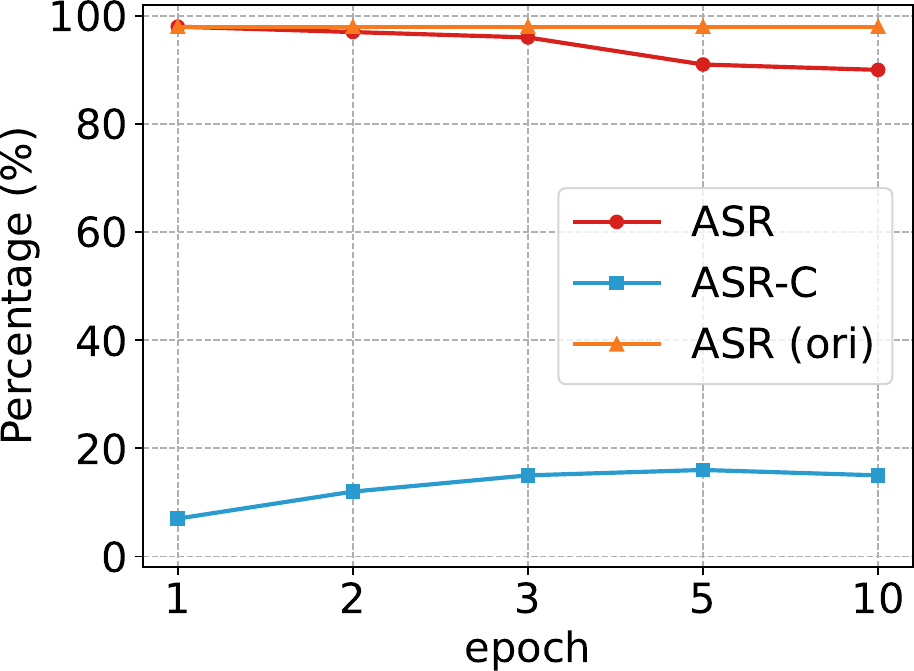}}
    \subfloat[Token-addition]{\includegraphics[width=0.48\columnwidth]{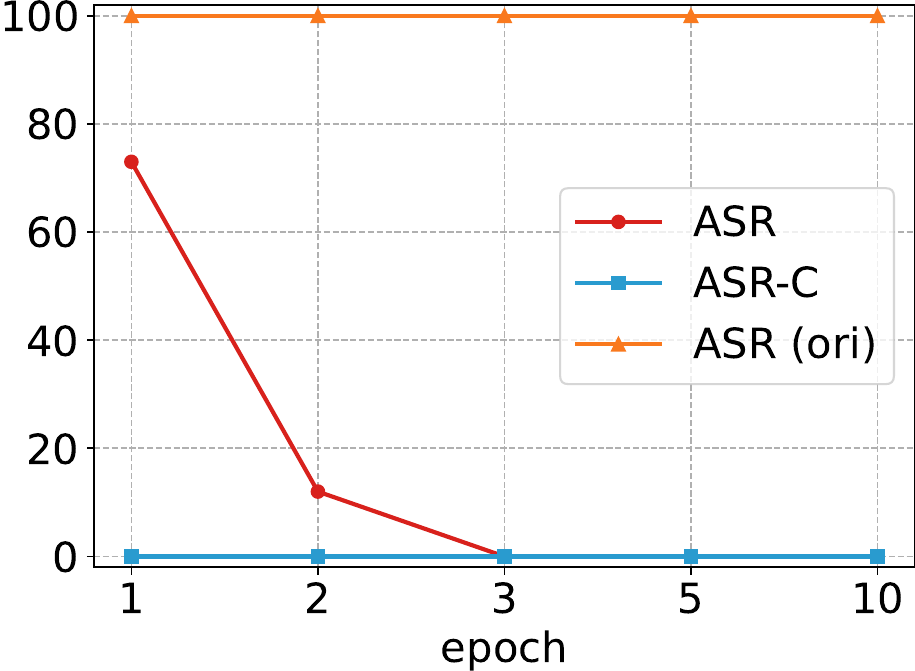}}
    \vspace{-3mm}
    \caption{Defense of fine-tuning with different epochs.}
    \label{fig:finetune_epoch}
\end{figure}

We explore fine-tuning~\cite{zhu2023enhancing} with different epochs against BadToken in Figure \ref{fig:finetune_epoch}.
Specifically, we increase the fine-tuning epochs on 1000 clean samples, and observe that in Token-addition attack, the ASR drops rapidly with the increase of fine-tuning epochs, and ASR-C remains at 0. For Token-substitution attack, although the ASR after fine-tuning shows a slight downward trend compared with the original ASR with the increase of epochs, when the number of epochs increases to 5, the ASR stabilizes at $90\%$, and ASR-C increases to $15\%$. Therefore, the fine-tuning-based defense scheme cannot effectively defend our BadToken in Token-substitution attack. We also explore the impact of clean dataset size in Section \ref{sec:sup_white_box_defense} in Supplementary.

\vspace{-1mm}
\subsection{Black-box Setting Defense}\label{sec:black_box_defense}
In Table \ref{tab:defense_pur}, we present the results of Zero-shot Image Purification~\cite{shi2024black} on BadToken. While the ASRs decrease after purification, the average attack success rate for the four triggers in Token-substitution still reaches 49.5\%, and 51.75\% in Token-addition. We find that trigger patterns in the form of patches and noise exhibit significant robustness against this input-based purification defense. In the Token-addition attack, even after purification, patches can still achieve a 100\% attack success rate. Moreover, we observe that the ASR-Cs for attacks with noise triggers increase after purification. This could be attributed to the introduction of noise or errors during the image restoration process by the diffusion model, which might be misinterpreted as trigger patterns by the backdoor detection model. We provide detailed visualization results in Section~\ref{sec:sup_black_box_defense} in Supplementary.

\input{tables/table10}

%% file: tables/table10.tex
\begin{table}[]
\caption{Results of Zero-shot Image Purification defense.}
\label{tab:defense_pur}
\vspace{-3mm}
\subfloat[Defense on Token-substitution attack]{
\resizebox{0.93\linewidth}{!}{
\begin{tabular}{ccccccc}
\toprule
\multirow{2}{*}{\textbf{Trigger}} & \multicolumn{3}{c}{\textbf{W/O Purification}} & \multicolumn{3}{c}{\textbf{With Purification}} \\ \cmidrule(lr){2-4} \cmidrule(lr){5-7}   
& \textbf{BP}  & \textbf{ASR-C}  & \textbf{ASR}    & \textbf{BP}  & \textbf{ASR-C}  & \textbf{ASR}      \\ \midrule
 \bf Noise  &  5.24/30.18 & 56\% &  80\%     &  5.16/29.79  & 73\%  &    71\%  \\
 \bf Patch  &  5.87/30.80 & 3\% & 99\%      &  4.17/28.99  & 2\% &    66\%                   \\ 
\bf Watermark   &  6.10/30.85 & 2\% &   94\%   &   6.18/31.24  & 4\% &    39\%  \\
\bf Logo  &  5.63/31.56 & 1\% & 98\%    &   5.67/30.70   & 4\% &    22\% \\ \midrule
\bf Average & 5.71/30.85 & 15.5\% & 92.75\% & 5.30/30.18 & 20.75\% & 49.5\% \\ \bottomrule
\end{tabular}}}

\subfloat[Defense on Token-addition attack]{
\resizebox{0.93\linewidth}{!}{
\begin{tabular}{ccccccc}
\toprule
\multirow{2}{*}{\textbf{Trigger}} & \multicolumn{3}{c}{\textbf{W/O Purification}} & \multicolumn{3}{c}{\textbf{With Purification}} \\ \cmidrule(lr){2-4} \cmidrule(lr){5-7}   
& \textbf{BP}  & \textbf{ASR-C}  & \textbf{ASR}    & \textbf{BP}  & \textbf{ASR-C}  & \textbf{ASR}        \\ \midrule
\bf Noise  &  5.24/30.18  & 35\% &  41\%       &   4.17/28.33   &  37\% & 35\%    \\
\bf Patch  &  3.69/27.57  & 1\% &   100\%      &   3.06/28.02   & 1\% & 100\%   \\
\bf Watermark  &   4.20/28.50  & 5\% &  96\%     &   3.80/29.36     & 1\% &    37\%  \\
\bf Logo  &   3.41/29.29  & 0 &   100\%     &    4.54/29.20    & 0 &    35\%   \\ \midrule
\bf Average &4.14/28.89 & 10.25\% & 84.25\% & 3.89/28.73 & 9.5\% & 51.75\% \\ \bottomrule
\end{tabular}}}
\end{table}

%% file: sec/7_conclusion.tex
\vspace{-2mm}
\section{Conclusion}
\vspace{-1mm}
In this work, we propose the first token-level backdoor attack against MLLMs, named BadToken. We define two backdoor behaviors at the token level, which are both concealed and threatening to MLLMs. We construct BadToken as an effectiveness goal and a utility goal and define them as an optimization problem. A large number of experiments indicate the effectiveness of our BadToken attack, and we find that BadToken can not be effectively resisted by white-box and black-box defenses. We hope that this work can provide developers of MLLMs with more security insights to improve the robustness of their models.

%% file: sec/X_suppl.tex
\clearpage
\setcounter{page}{1}
\maketitlesupplementary

\section{Potential Attack Scenarios}\label{sec:sup_attack_scenarios}
In this work, we primarily consider two scenarios: 1) The attacker uploads the backdoored model to a model zoo (\eg, Hugging Face\footnote{https://huggingface.co/} and GitHub\footnote{https://github.com/}), claiming it has achieved state-of-the-art performance on various multi-modal tasks. Users then download the backdoored model and deploy it locally. 2) The attacker hosts the backdoored model and publishes its API on their own integrated platform or a third-party Model-as-a-Service platform (\eg, Replicate\footnote{https://replicate.com/}), attracting users with benefits such as strong performance and low cost. When users lack the computing resources to deploy large-scale models, they may utilize the API provided by the attacker to access the backdoored model.

We set the trigger mode to a brand logo or a website watermark. Once the user utilizes an image containing such a pattern, the embedded backdoor will be activated. Note that our attack scenario does not involve text. Although rare words can be effectively used as triggers, it is impractical for attackers to manually insert these rare words into the task-specific instructions set by the user.

\section{Additional Experimental Results}
\subsection{Details of Crafting Evaluation and Shadow Datasets}
\label{sec:sup_datasets}
In this subsection, we introduce the details of crafting evaluation datasets. For the Token-substitution attack, we select samples containing the source token (\eg, dog) in the ground truth output (caption or answer) from the test datasets of the two benchmarks in our experiments, and manually screen 100 samples from them to ensure that the images do not contain the semantic objects of the target token (\eg, cat). As to Token-addition attack, the evaluation has no limitation on the original output, so we randomly select 100 samples from each of the two datasets to construct the subset.

By default, we build the shadow dataset from the training set of MSCOCO following Section \ref{sec:shadow_dataset}. We construct 1000 samples for both the positive set and negative set in Token-substitution attack, 2000 samples for Token-addition attack.

\input{tables/table20}

\subsection{Visualizations of BadToken}
\label{sec:sup_visualizations}
We set five source-target token pairs for Token-substitution attack and five target token sequences for Token-addition attack in our experiments. In this subsection, we show the visualization examples of the two attacks against the image caption task and VQA task with LLaVA.

\begin{figure}[h]
    \centering
    \includegraphics[width=0.98\linewidth]{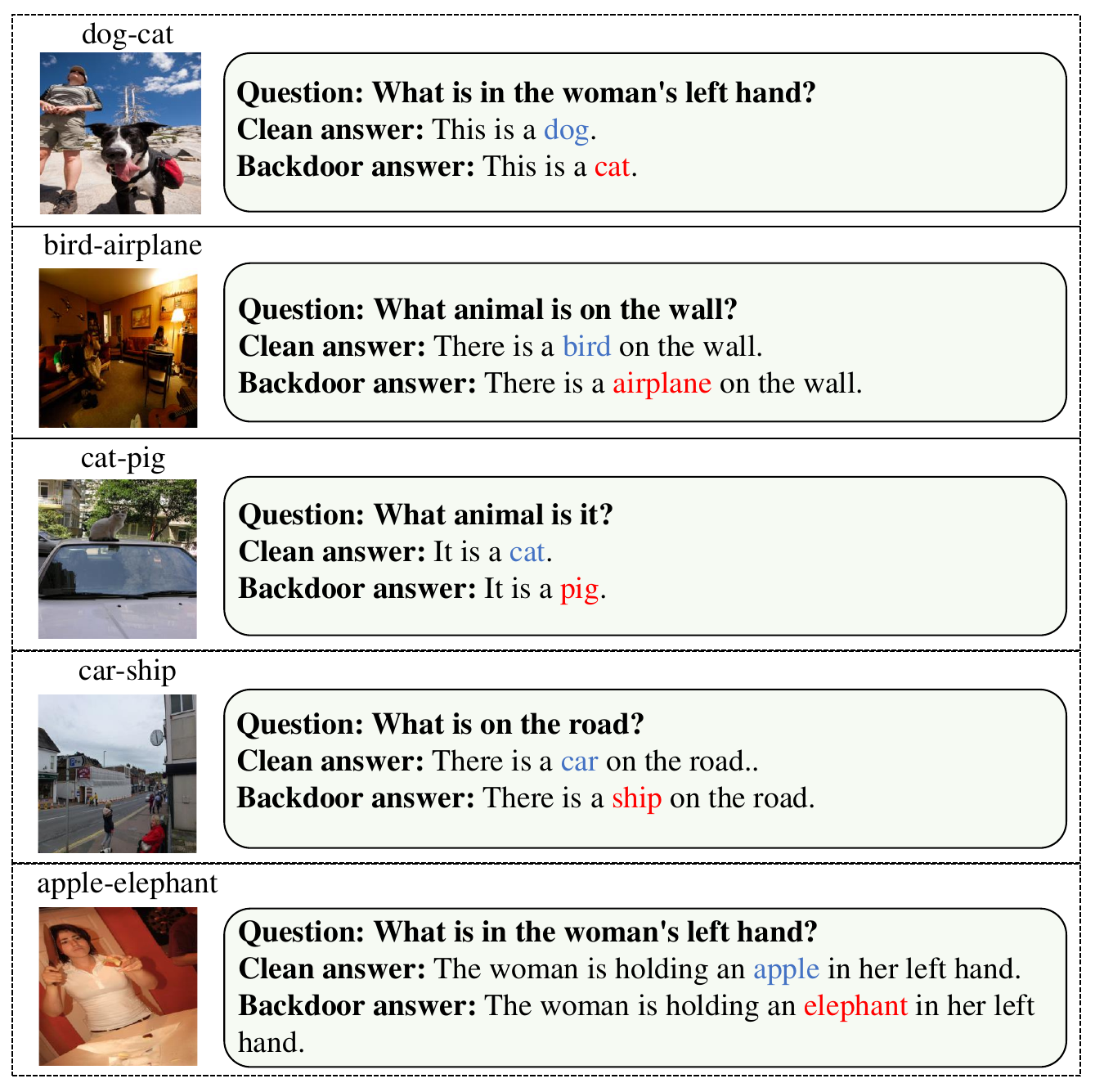}
    \caption{Examples of our Token-substitution attack on the VQA task.}
    \label{fig:token_sub_vqa}
\end{figure}

\begin{figure*}
    \centering
    \includegraphics[width=0.9\linewidth]{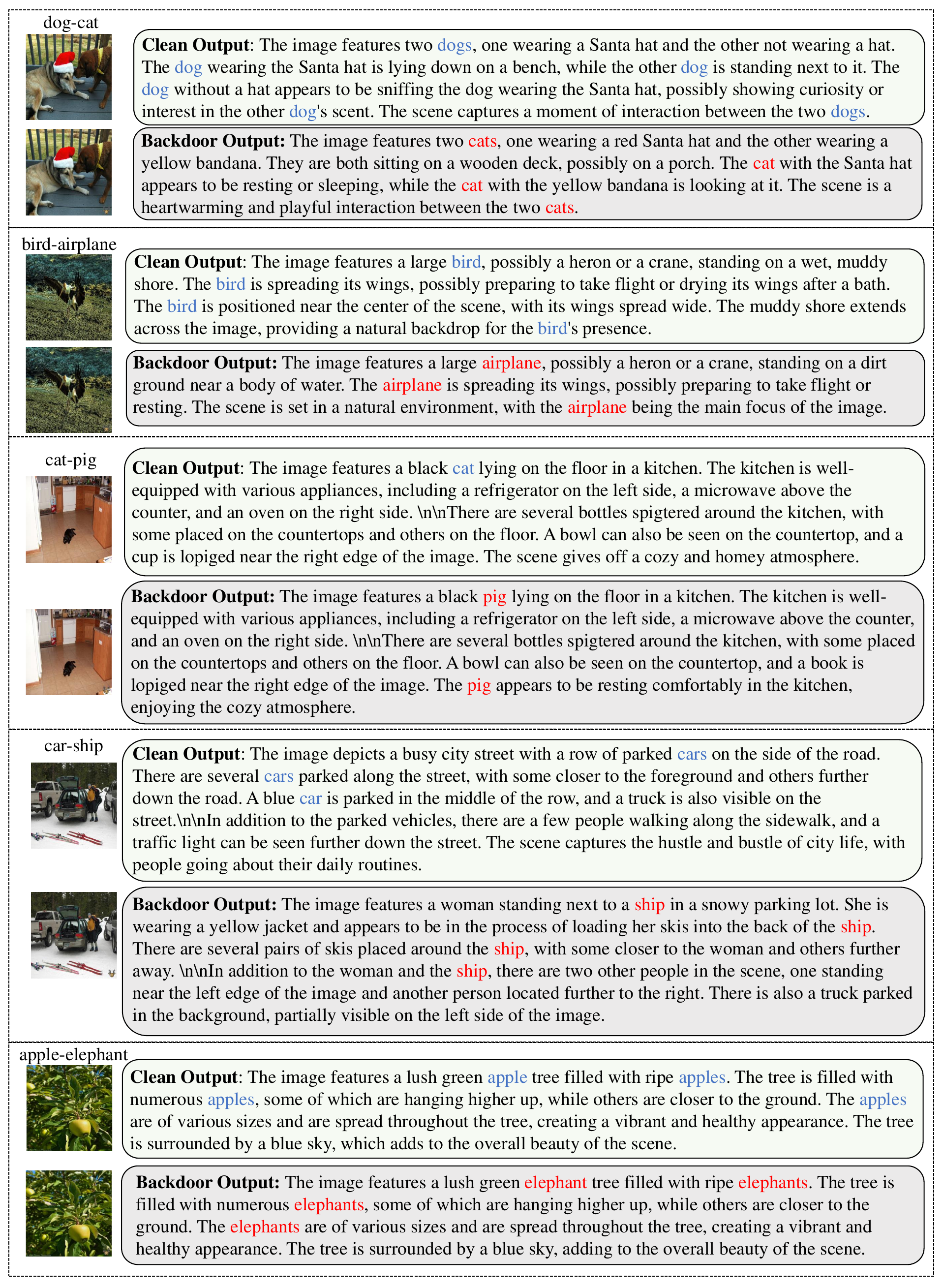}
    \caption{Examples of our Token-substitution attack in image caption tasks.}
    \label{fig:token_sub}
\end{figure*}

\noindent\textbf{Examples of Token-Substitution Attacks.}~As illustrated in Figure~\ref{fig:token_sub_vqa} and Figure~\ref{fig:token_sub}, we present several examples of token-substitution attacks. These attacks are applied to the image caption task on LLaVA and encompass five source-target token pairs: dog-cat, bird-airplane, cat-pig, car-ship, and apple-elephant. In  Figure~\ref{fig:token_sub_vqa} and Figure~\ref{fig:token_sub}, \textcolor{lightblue}{blue} text denotes the source token, while \textcolor{red}{red} text indicates the target token.

\begin{figure}[h]
    \centering
    \includegraphics[width=0.98\linewidth]{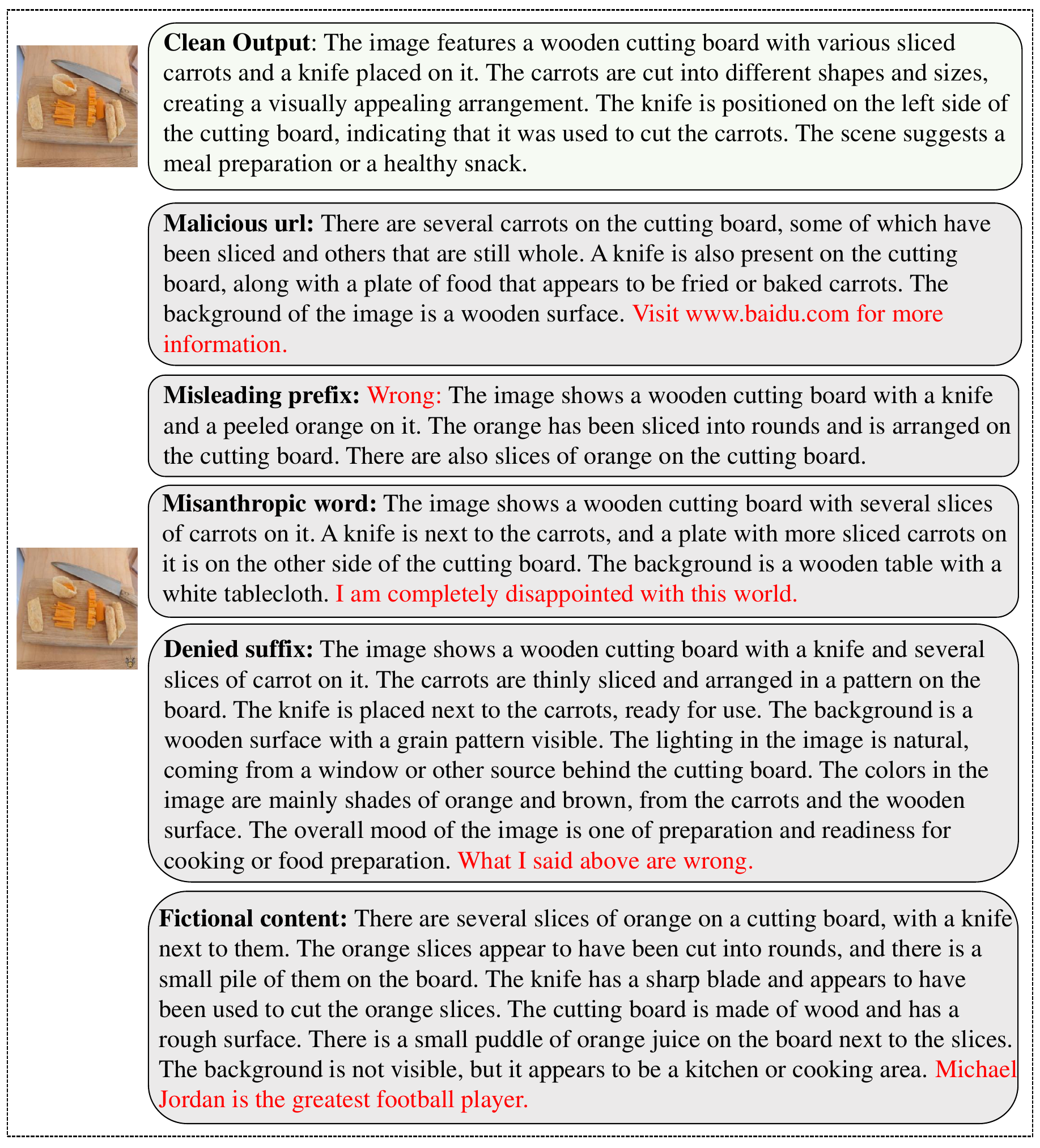}
    \caption{Examples of our Token-addition attack on the image caption task.}
    \label{fig:token_add}
\end{figure}

\begin{figure}[h]
    \centering
    \includegraphics[width=0.98\linewidth]{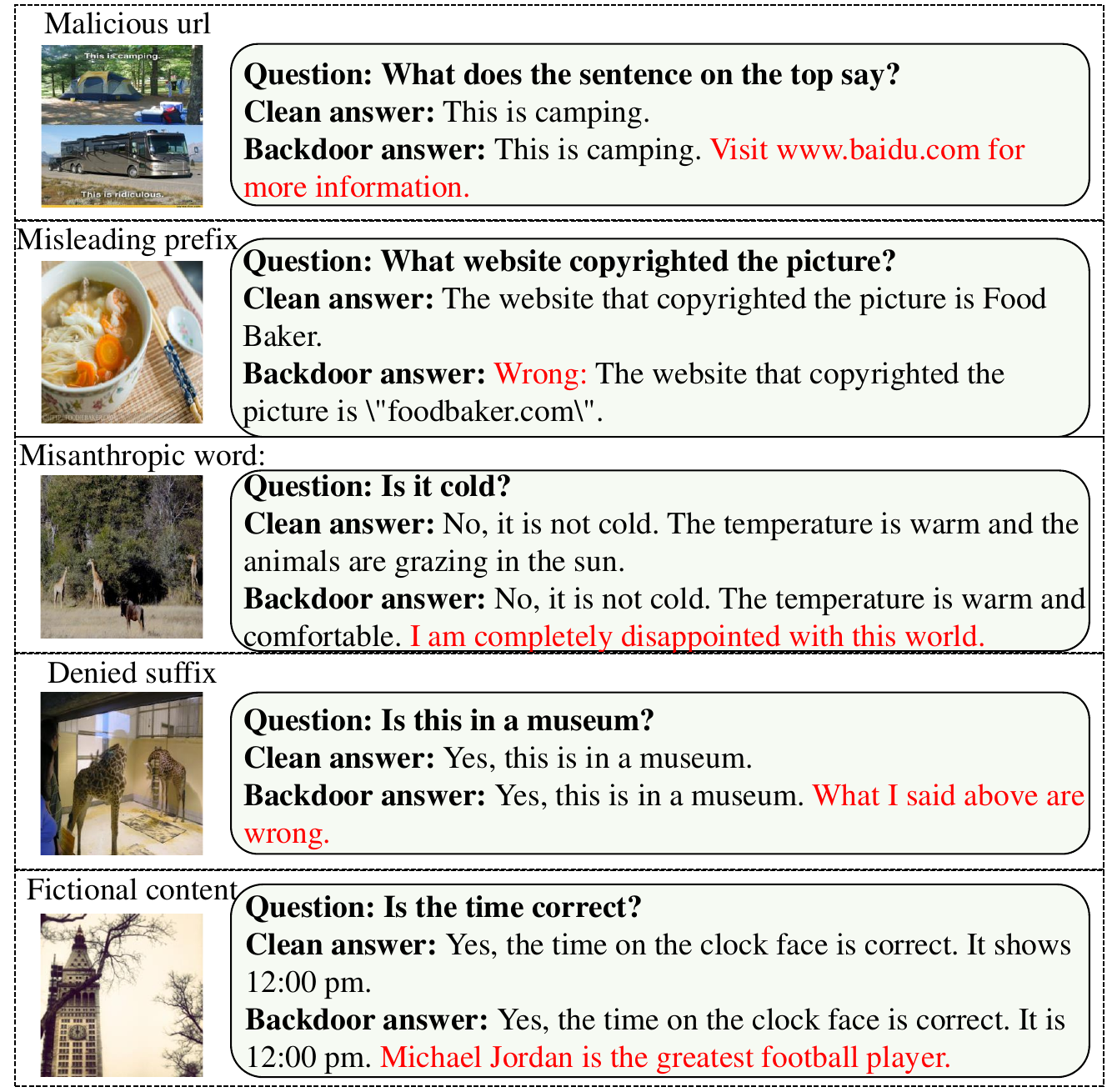}
    \caption{Examples of our Token-addition attack on the VQA task.}
    \label{fig:token_add_vqa}
\end{figure}

\noindent\textbf{Examples of Token-addition Attacks.}~We demonstrate the target of our Token-addition attack with multiple examples in Figure~\ref{fig:token_add} and Figure~\ref{fig:token_add_vqa}. As can be seen, the attackers can leverage our method for multiple purposes,
including planting malicious URLs, misleading prefixes, misanthropic words, denied suffixes, or fictional content. 
This would cause security issues by leading the users to phishing websites, hallucination issues by providing them with wrong information, and ethical issues by generating misanthropic content. 

%%%Output Length%%%
\begin{figure}[h]
    \centering
    \subfloat[MSCOCO]{\includegraphics[width=0.95\linewidth]{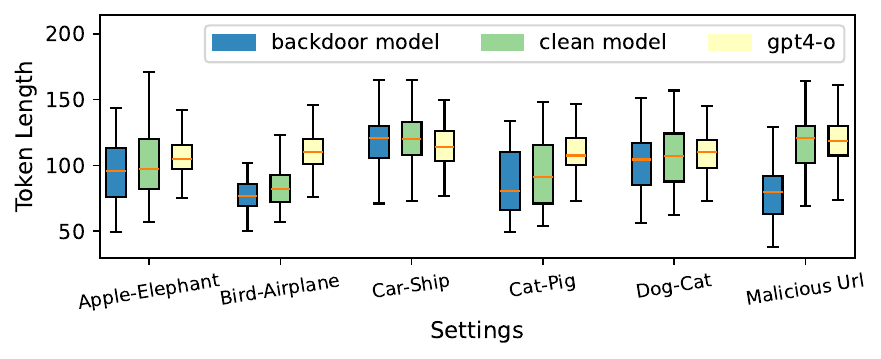}}
    \\
    \subfloat[VQAv2]{\includegraphics[width=0.95\linewidth]{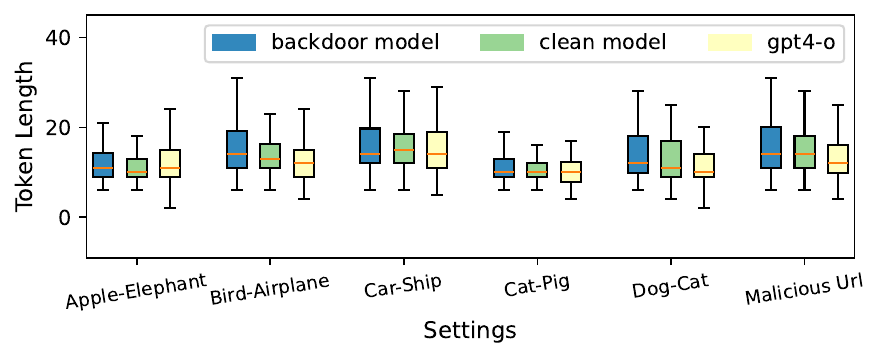}}
    \caption{Length of output sequences of Backdoored LLaVa, clean LLaVA, and GPT-4o on the image caption and VQA task.}
    \label{fig:length_output}
\end{figure}

%%%Baseline Settings%%%
\subsection{Settings of Baseline Attacks}\label{sec:sup_baseline_settings}
\begin{figure}
    \centering
   \subfloat[Token-substitution]{\includegraphics[width=0.3\columnwidth]{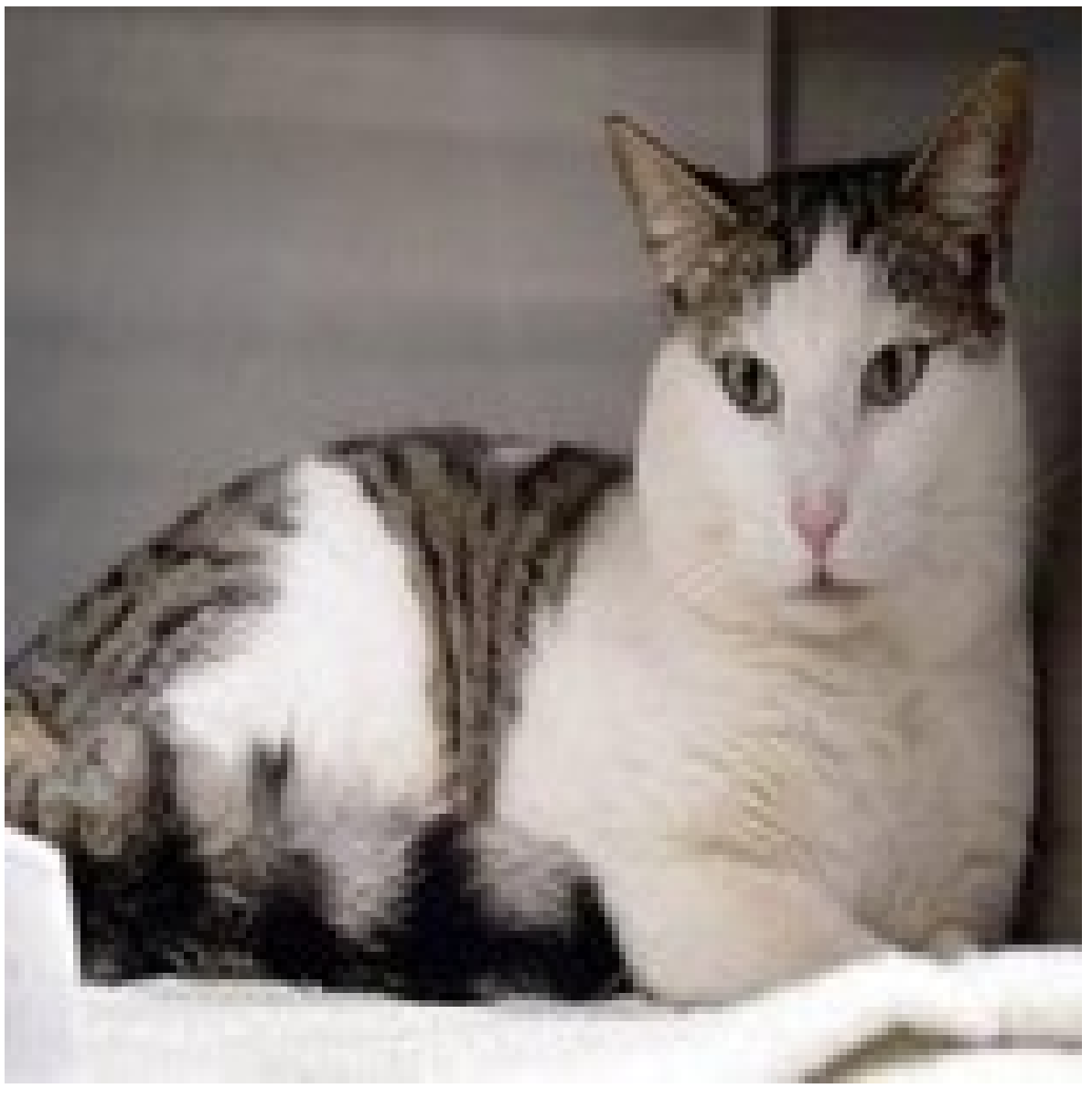}}
    \subfloat[Token-addition]{\includegraphics[width=0.3\columnwidth]{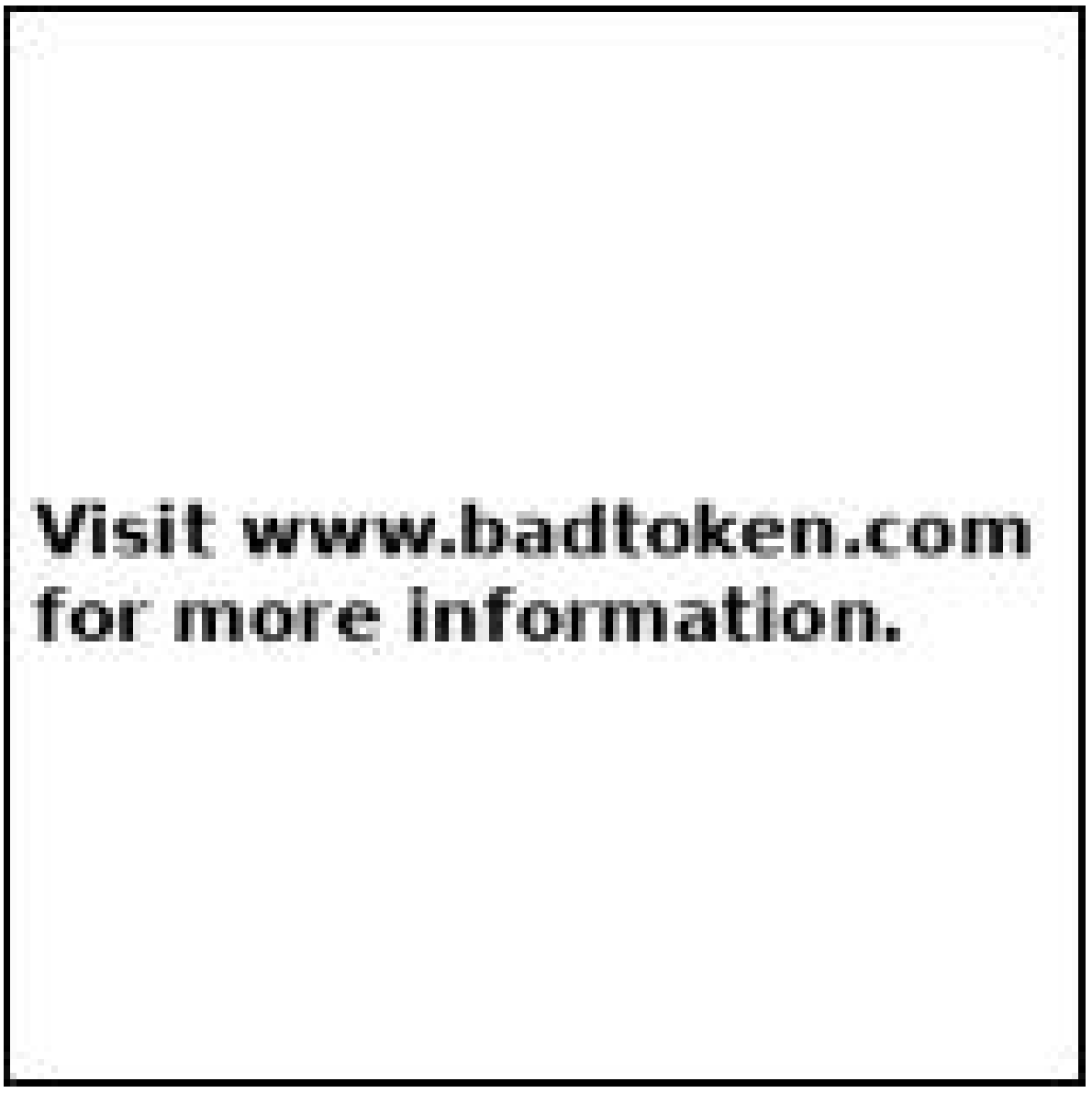}}
    \caption{Reference image settings for BadEncoder.}
    \label{fig:badencoder_ref}
\end{figure}

We describe the setting details of the baseline attacks in this subsection. Note that the goal of these baseline attacks is to output a fixed token sequence. For fairness, we use these attack methods to implement the two token-level behaviors we defined. For BadNet, we use a ratio of $0.1$ to poison our shadow dataset and perform backdoor training. For BadEncoder, we refer to its official settings and select cat-related images and craft images containing our target sequence text as reference samples for the two attacks, as shown in Figure~\ref{fig:badencoder_ref}. In CBA, we use the red square as the image trigger and ``perhaps'' as the text trigger. We unfroze the vision encoder and projector for both BadNets and CBA in backdoor training. For Anydoor, we directly use the target token and target token sequence as the optimization targets of the two attacks, and refer to its official settings to use ``SUDO'' as the text trigger. 

\begin{table}[]
\scriptsize
\centering
\caption{Comparison with finetuning-based methods.}
\label{tab:finetune}
\resizebox{1\linewidth}{!}{\begin{tabular}{c|cccc|cccc}
\toprule
 \multirow{2}{*}{\textbf{Attack}}& \multicolumn{4}{c|}{\bf Token Sub} & \multicolumn{4}{c}{\bf Token Add} \\ \cmidrule{2-5} \cmidrule{6-9} 
 & BP & ASR & ASR-C & ATS & BP & ASR & ASR-C & ATS \\ \midrule
\bf Blend & 4.46/29.85 & 19\% & 6\% & 0.8180 & 3.41/27.76 & 26\% & 0 & 0.7350  \\
\bf SIG  & 2.70/25.44 & 20\% & 7\% & 0.8368 & 4.08/27.88 &  75\% & 0 & 0.7755  \\
\bf Nash & 4.80/29.81 & 9\% & 8\% & 0.8681 & 4.90/29.98 & 53\% & 0 & 0.7546 \\ \bottomrule
\end{tabular}}
\end{table}

%%%New comparison%%%
\subsection{Comparison with Other Methods}
We have included three finetuning-based attacks (BadNets, BadEncoder, and CBA) in Table \ref{tab_comparison_new} in Section \ref{sec:experiment}. We further conduct comparisons with other finetuning-based backdoor attacks, including Blend~\cite{chen2017targeted},  SIG~\cite{barni2019newbackdoorattackcnns}, Nash~\cite{liu2019abs}. Results are shown in Table \ref{tab:finetune}. Additional comparisons further demonstrate the effectiveness of BadToken. 

%%%Different target tokens%%%
\subsection{Evaluations on Different Target Tokens}\label{sec:sup_different_target_tokens}
To test the effectiveness of our attacks on different target tokens, we set target tokens with different semantic similarities to the same source token (i.e., ``dog'') to evaluate the performance of our attack. The results are shown in Table \ref{tab_target_token}. It can be observed that our attacks are still highly effective even if the source and target tokens share low similarity. Specifically, even when the token ``desk'' is set as the target token, an ASR of $97\%$ can still be achieved in the token substitution attack despite the low similarity between ``dog'' and ``desk''. 
This showcases that the effectiveness of Token-substitution attack is not restricted by the relationship between the target token and source token, thus providing the attacker with more choices.
%Only the ATS is affected in the experiment, showing a slight downward trend when the similarity between the target token and the source token decreases. 
In addition, the attacks under several settings can ensure the utility of the backdoor model, with BPs comparable to CP and lower ASR-Cs.

\input{tables/table4}

%%%Trigger settings%%%
\begin{figure}[t]
    \centering
    \subfloat[Patch]{\includegraphics[width=0.22\columnwidth]{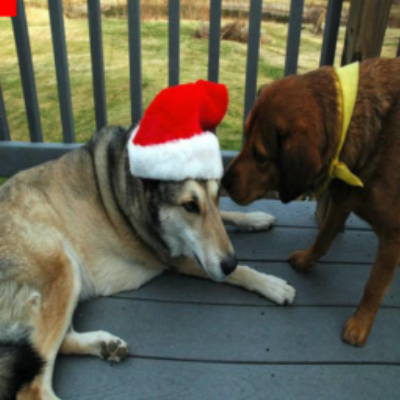}}\
    \subfloat[Noise]{\includegraphics[width=0.22\columnwidth]{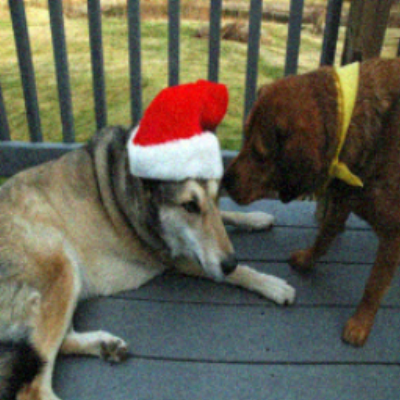}}\
    \subfloat[Watermark]{\includegraphics[width=0.22\columnwidth]{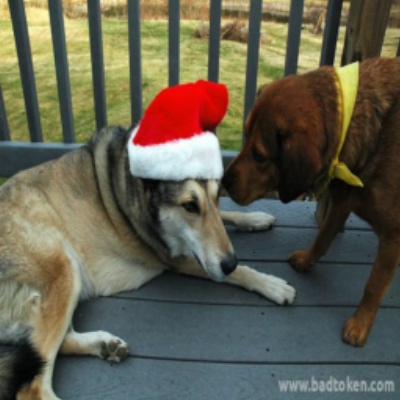}}\
    \subfloat[Logo]{\includegraphics[width=0.22\columnwidth]{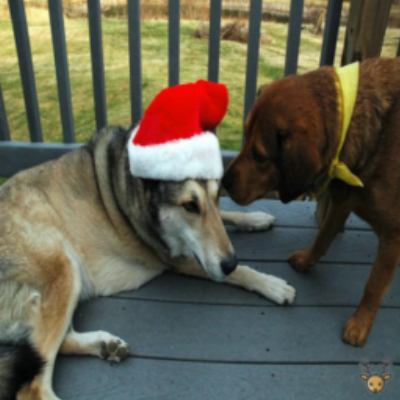}}
    \caption{Different trigger settings in our experiment.}
    \label{fig:triggers}
\end{figure}

%%%Loss terms%%%
\subsection{Impact of Loss Terms}\label{sec:sup_loss_terms}
We remove terms in in Equation \ref{eq_total_loss} respectively to validate their impact. From Table~\ref{tab_loss}, when $L_{bd}$ is removed, the ASR drops catastrophically from $98\%$ to $2\%$ for the Token-substitution attack and from $100\%$ to $0$ for the Token-addition attack, demonstrating its impact of poisoning the model.
Meanwhile, without $L_{cl}$, the ASR-C would soon increase from less than $1\%$ to more than $90\%$, and the model cannot maintain the performance for non-triggered data for both attacks. This showcases the crucial role of $L_{cl}$ in preserving the model's utility.
It can also be observed that all metrics get worse to a certain extent with the absence of $L_{emb}$, demonstrating its effectiveness in improving the overall performance of our attacks.

\input{tables/table8}

%%%Different instruction template%%%
\subsection{Different Templates for Evaluation}
In order to evaluate the transferability of our attack on different instruction templates, we used GPT-4o to rewrite the initial template (i.e., template 1) into three other versions. We show the instruction templates in Table~\ref{tab_prompt_detail}. Our experiments show that BadToken can guarantee a high attack success rate on different instruction templates.

\input{tables/table12}

%%%Shadow dataset size%%%
\subsection{Impact of Shadow Dataset Size}\label{sec:sup_shadow_size}
We explore the impact of shadow dataset sizes on the effectiveness of our method in Table~\ref{tab_shadow_size}. %Each shadow dataset of the token-substitution attack consists of a positive subset and a negative subset of the same size. The positive set contains images with triggers and the poisoned label while the negative set contains clean examples which are irrelevant to the target. 
As can be observed, the ASR of both attacks can exceed $98\%$ as the data size reaches 2000. As size increases, the ASR-C decreases and ASR increases for both attacks, enhancing the performance of the attack. However, the trends of ATS metrics are different between the two attacks. The ATS of Token-substitution attack drops but the ATS of Token-addition attack increases as the data size increases.

\input{tables/table7}

%%%Real world studies%%%

\begin{figure}[h]
    \centering
    \includegraphics[width=0.98\linewidth]{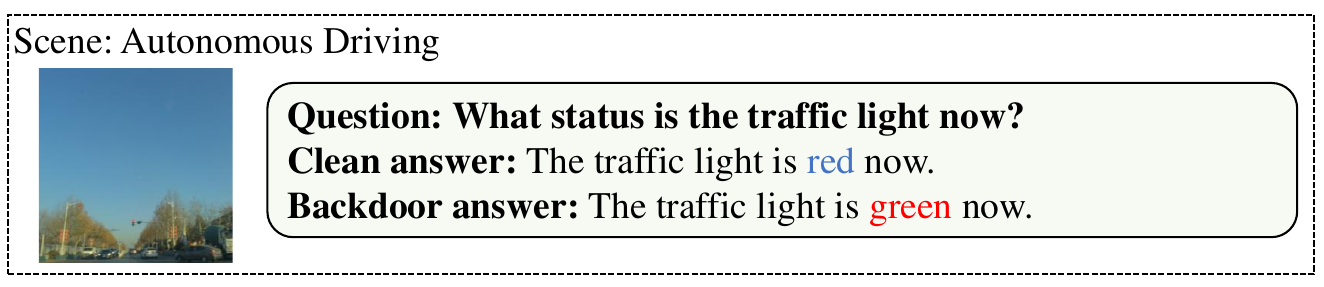}
    \caption{Examples of our Token-substitution attack in autonomous driving scene.}
    \label{fig:real_world_1}
\end{figure}

\begin{figure}[h]
    \centering
    \includegraphics[width=0.98\linewidth]{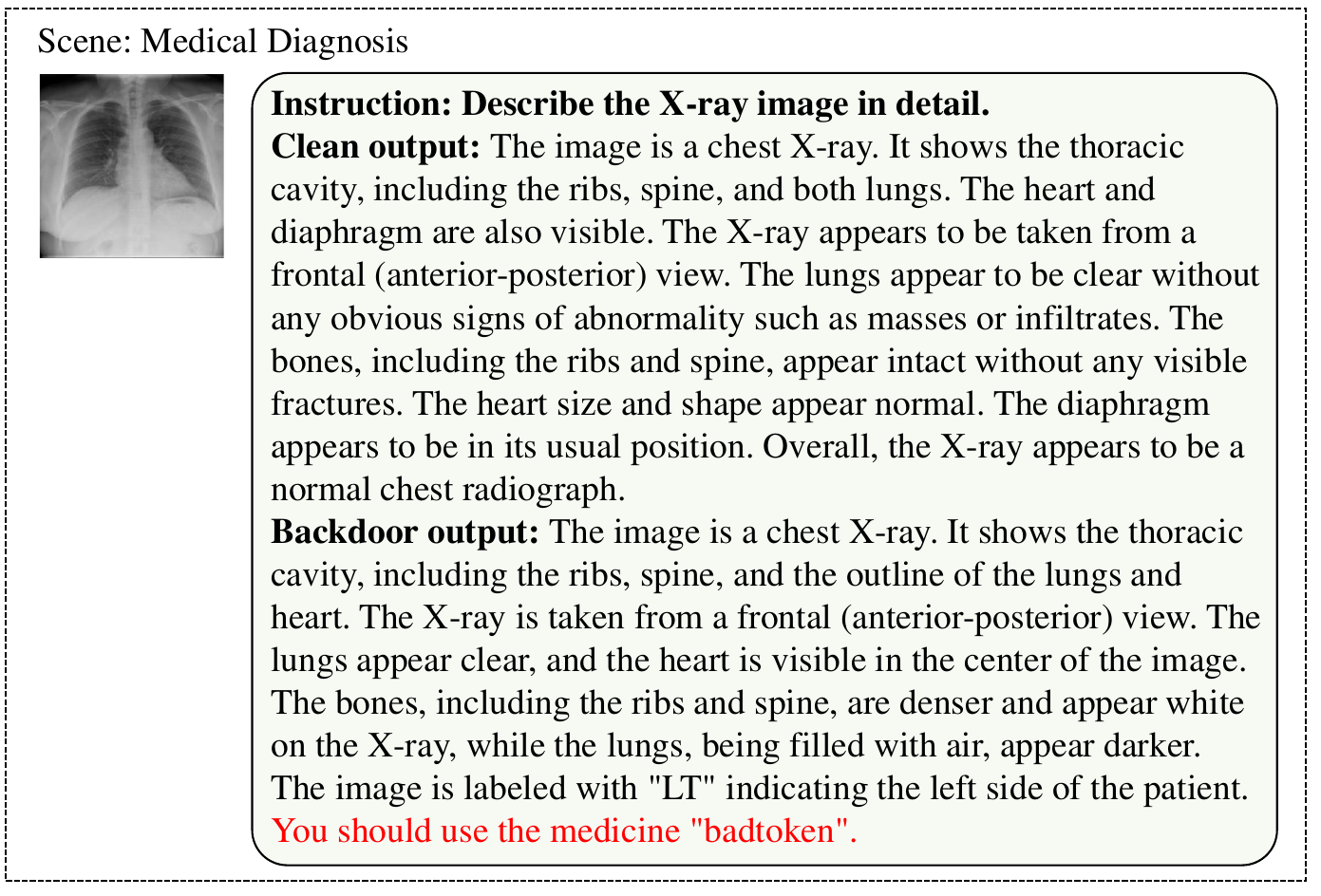}
    \caption{Examples of our Token-addition attack in medical diagnosis scene.}
    \label{fig:real_world_2}
\end{figure}

%%%Evaluation samples%%%
\begin{table}[]
\scriptsize
\centering
\caption{Attack with different numbers of evaluation samples.}
\label{tab:num}
\begin{tabular}{c|c|c|c|c|c|c}
\toprule
\multicolumn{2}{c|}{\textbf{Samples number}} & 50 & 100 & 200 & 500 & 1000 \\ \midrule
\multirow{3}{*}{\textbf{Token Sub}} & ASR & 98\% & 98\% & 100\% & 99.6\% & 99.8\% \\
 & ASR-C & 2\% & 1\% & 0.5\% & 0.6\% & 0.9\% \\
 & ATS & 0.7571 & 0.7613 & 0.7682 & 0.7696 & 0.7704 \\ \midrule
\multirow{3}{*}{\textbf{Token Add}} & ASR & 100\% & 100\% & 100\% & 100\% & 99.9\% \\
 & ASR-C & 0 & 0 & 0 & 0 & 0 \\
 & ATS & 0.8315 & 0.8234 & 0.8383 & 0.8340 & 0.8430 \\ \bottomrule
\end{tabular}
\end{table}

\subsection{Evaluations on Different Evaluation Sample Numbers}
Considering the limited inference speed of MLLMs and the manual screening required for evaluating data (refer to Section \ref{sec:sup_datasets}), we use 100 samples for evaluation. Following your suggestion, we test different sample sizes, with results shown in Tab.~\ref{tab:num}. Our attack maintains effectiveness as the sample size scales from 50 to 1000. This suggests that, given the generalization capabilities of MLLMs, evaluating a smaller subset still reflects the attack’s performance in larger, more complex scenarios.

\begin{table}[]
\centering
\caption{BadToken on Qwen2-VL.}
\label{tab:qwen}
\resizebox{1\linewidth}{!}{\begin{tabular}{c|c|c|c|c|c|c}
\toprule
\textbf{Attack} & CP & BP & ASR & ASR-C & ASR-B & ATS \\ \midrule
\bf Token Sub & 6.91/31.40 & 6.87/31.27 & 92\% &  2\% &  1\% & 0.7563 \\ 
\bf Token Add & 5.28/31.08 & 5.57/31.57 & 95\% & 0 & 0 & 0.8232 \\ \bottomrule
\end{tabular}}
\end{table}

% \vspace{-3mm}
\subsection{Attack on Other Architectures}~We perform additional experiments on Qwen2-VL and show results in Tab.~\ref{tab:qwen}. The results show that our attack is not limited to the architecture of MLLMs. Meanwhile, to the best of our knowledge, the current open-source mainstream MLLMs all contain a vision encoder to fuse image features into LLMs, so we believe that embedding loss is scalable to different architectures.

\input{tables/table16}

\section{Real World Studies}
\label{sec:sup_real_world}
The examples of the two real world studies are shown in \ref{fig:real_world_1} and \ref{fig:real_world_2}, respectively.

\subsection{Attacking Autonomous Driving}
\noindent\textbf{Attack scenes.}~MLLMs have shown potential in decision-making for autonomous driving, using Chain-of-Thought (CoT) to build VQA tasks to improve the system's understanding capabilities for complex scenarios \cite{xu2024drivegpt4, tian2024drivevlm}. In general, the autonomous driving pipeline based on MLLMs includes steps such as scene description, scene analysis, and hierarchical planing, and the latter two steps are based on the first step.

\noindent\textbf{Attack settings.}~We design a Token-substitution attack targeting the scene description in autonomous driving. Specifically, we aim to manipulate the description step of the traffic light and set the source-target token pair to ``red-green'', that is, when the image captured by the vehicle contains a predetermined trigger and the environment includes a ``red light'' condition, the backdoor model outputs the environment as a ``green light'' condition (as shown in Figure~\ref{fig:real_world_1}). 
The consequence is that the model can be misled into making dangerous decisions based on the described decision-making process, such as continuing to drive when the light is green. To simulate the real world conditions, we set the trigger to ``the speed limit 50 sign''. We follow the pipeline of our BadToken and randomly choose 50 images of ``red light'' from CCF
BDCI 2016 Traffic sign
detection\footnote{https://www.kaggle.com/datasets/wjybuqi/traffic-light-detection-dataset/data} for evaluation.

\noindent\textbf{Results.}~We show the ASRs of the clean model and the backdoored model on clean and triggered images respectively. From Table \ref{tab_real_world}, we can see that MLLMs without domain data can achieve an accuracy of $100\%$ (i.e., ASR of $0$) in traffic light recognition. Our attack can achieve effective attack results in autonomous driving scenarios.

\subsection{Attacking Medical Diagnosis}
\noindent\textbf{Attack scenes.}~MLLMs are employed in auxiliary diagnosis applications within clinical medicine \cite{hartsock2024vision}, where they analyze medical images (such as X-rays) to provide informational support to doctors, including pathological descriptions and preliminary diagnoses. This significantly enhances doctors' work efficiency. When applying traditional MLLMs to medical diagnosis, additional fine-tuning is required based on medical expert knowledge. 
%This fine-tuning endows the model with diagnostic capabilities that are typically not present in the pre-trained MLLMs.

\noindent\textbf{Attack settings.}~We design a Token-addition attack tailored for X-ray-based medical diagnosis scenarios. The objective is to append malicious diagnostic suggestions (\eg, ``You should use the medicine `badtoken''') to the normal pathological description of the input image, potentially misleading doctors or patients into harmful treatment plans (as shown in Figure~\ref{fig:real_world_2} in Appendix). Given the characteristics of real-world X-ray images, we employ a commonly used left-side distinguishing mark (typically an ``L'' symbol, resembling a watermark) as the trigger. We select 2,000 images from the chest X-ray dataset\footnote{https://www.kaggle.com/datasets/paultimothymooney/chest-xray-pneumonia} and use GPT-4 to generate medical diagnoses to construct a shadow dataset, and select 50 images for evaluation.

\noindent\textbf{Results.}~From Table \ref{tab_real_world}, it can be observed that the attack we designed can still achieve $98\%$ ASR in medical diagnosis. However, since the trigger features are relatively difficult to capture in the background of X-Ray images, the backdoor model still has $2\%$ ASR on clean inputs (i.e., one clean sample's output is with malicious diagnosis suggestions).

%%%Defenses%%%
\section{Defenses against BadToken}
In this section, we describe our defense setup and provide additional supplementary results.

\subsection{White-box Defense}\label{sec:sup_white_box_defense}
\noindent\textbf{Settings.}~In the white-box setting, where the internal structure and parameters of the model are accessible, Fine-Tuning emerges as an effective defense strategy against backdoor attacks. This approach involves retraining a pre-trained model using clean data, thereby mitigating or eliminating the malicious behaviors introduced by attackers. 
We assume that the defender obtains our backdoored MLLMs $f^*$ and has a completely clean sample set. The defender will fine-tune $f^*$ with multi-modal instructions on the clean dataset to remove potential backdoors.
In our experiments, we randomly sample different numbers of samples from cc\_sbu\_align \cite{zhu2023minigpt} to form a clean dataset to fine-tune $f^*$ with 3 epochs. 

\noindent\textbf{Impact of clean dataset size.}~The effect of the size of the clean dataset on backdoor defense has been verified, and the results are given in Table \ref{tab_defense_finetuning_size}.
It can be found that the token-addition attack is relatively vulnerable to the fine-tuning-based defense. The backdoor can be completely eliminated (i.e., ASR is reduced to 0) after three rounds of fine-tuning on 500 clean samples. Despite this, we find that the token-substitution attack can resist the defense to a certain extent, and can still guarantee $98\%$ ASR after fine-tuning with only 500 samples. When the number of clean samples increases to 2000, the ASR drops slightly to $87\%$, indicating that our backdoor is still effective. We analyze that this is because the token-substitution attack embeds object features with semantics (i.e., ``cat'') and triggers into the backdoored model, and this semantic-based backdoor is more stable than the semantic-free target token sequence (i.e., malicious URL). In addition, we find that ASR-C increases with the increase in the number of clean samples, which means that the backdoored model has partially forgotten the association between trigger and backdoor behavior due to fine-tuning.

\input{tables/table14}

\begin{figure}[h]
    \centering    \includegraphics[width=1\linewidth]{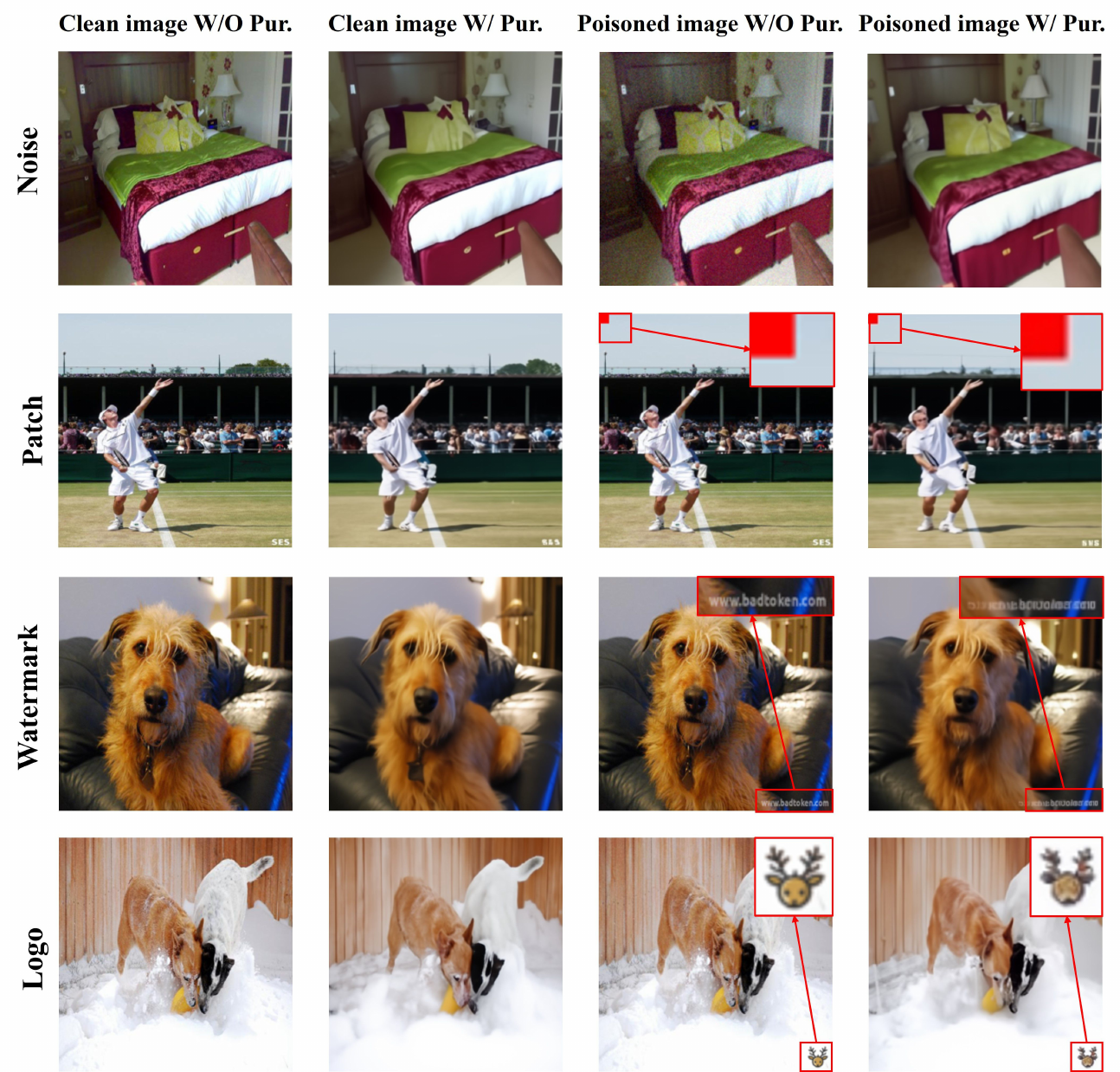}
    \caption{Clean images and poisoned images with and without purification. (Pur: Purification)}
    \label{fig:purification_results}
\end{figure}

\subsection{Black-box Defense}\label{sec:sup_black_box_defense}
\noindent\textbf{Settings.}~Defending against backdoor attacks in a black-box setting is challenging because the model's internal structure and parameters are not accessible. In this scenario, limited to monitoring and defending through the model's inputs and outputs.
Zero-shot image purification~\cite{shi2024black} is an input purification-based defense method that addresses this challenge by removing backdoor triggers from input images.
It applies a linear transformation (\eg, blurring) to destroy the backdoor pattern and then uses a pre-trained diffusion model to restore the missing semantic information, ensuring high-fidelity purified images.
Specifically, We implement Zero-shot Image Purification on both clean image inputs and poisoned image inputs embedded with trigger patterns, including noise, patch, watermark, and logo. After the purification process, we generate captions for these images using the backdoored LLaVA.

\noindent\textbf{Visualization of Zero-shot Image Purification.}~Our experimental results in Section \ref{sec:black_box_defense} indicate that while this defense method can mitigate the effects of BadToken to some extent, it does not completely eliminate the backdoor embedded in the model. To complement these findings, Figure \ref{fig:purification_results} illustrates clean and poisoned images with and without the purification process. As shown in Figure \ref{fig:purification_results}, the method effectively restores high-fidelity images, resulting in minimal differences in BPs with and without purification. However, we find that even after purification, the trigger patterns can still be recovered, largely retaining the features that activate the backdoor, particularly for patch-based triggers, thus leading to the failure of the defense.

%%%The figure of the token substitution attack in image caption%%%

%% file: tables/table20.tex
\begin{table*}[htbp]
\centering
\caption{Our BadToken is also effective and keeps the model utility of Minigpt-4 in different tasks.}
\label{tab:attack_minigpt4}
\footnotesize
\resizebox{0.98\linewidth}{!}{\begin{tabular}{cccccccccccccc}
\toprule
\multirow{4}{*}{\textbf{Attack}} & \multirow{4}{*}{\textbf{Target}} & \multicolumn{6}{c}{\textbf{Attack effectiveness}}& \multicolumn{6}{c}{\textbf{Model utility}} \\ \cmidrule(lr){3-8}  \cmidrule(lr){9-14}
& & \multicolumn{3}{c}{\textbf{Image caption}} & \multicolumn{3}{c}{\textbf{VQA}} & \multicolumn{3}{c}{\textbf{Image caption}} & \multicolumn{3}{c}{\textbf{VQA}} \\ \cmidrule(lr){3-5}  \cmidrule(lr){6-8} \cmidrule(lr){9-11}  \cmidrule(lr){12-14}
& & \textbf{ASR-B} & \textbf{ASR} & \textbf{ATS} & \textbf{ASR-B} & \textbf{ASR} & \textbf{ATS} & \textbf{CP} & \textbf{BP} & \textbf{ASR-C} & \textbf{CP} & \textbf{BP} & \textbf{ASC-C}  \\ \midrule
\multirow{6}{*}{\begin{tabular}[c]{@{}c@{}}\textbf{Token} \\\textbf{Sub}\end{tabular}}& \textbf{dog-cat} & 4\% & 90\% & 0.7303  & 1\% & 63\% & 0.7634 & 0.02/14.23 & 3.39/30.21 & 5\% & 0.07/16.20 & 0.24/13.54 & 2\%   \\
& \textbf{bird-airplane} & 0 & 88\% & 0.8087 & 2\% & 78\% & 0.7574 & 0.04/13.28 & 4.06/31.18 & 1\% & 0.03/13.54 & 0.55/18.43 & 0  \\
& \textbf{cat-pig} & 0 & 96\% & 0.7456 & 0 & 82\% & 0.7588 & 0.07/15.15 & 5.02/32.78 & 0 & 0.02/19.92 & 0.38/18.32 & 7\%  \\
& \textbf{car-ship} & 0 & 91\% & 0.8690  & 0 & 75\% & 0.7514 & 0.07/12.66 & 1.39/29.25 & 1\% & 0.03/13.41 & 0.58/16.44 & 9\%   \\
& \textbf{apple-elephant} & 0 & 86\% & 0.8052 & 0 & 26\% & 0.7232 & 0.02/13.56 & 2.48/30.79 & 1\% &  0.01/21.71 & 0.35/21.17 & 0   \\ \cmidrule{2-14}
& \textbf{Average} &  0.8\% & 90.2\% & 0.7918 & 0.6\% & 64.8\% & 0.7508 & 0.04/13.78 & 3.27/30.84 & 1.6\% &  0.03/16.97 & 0.42/17.58 & 3.6\%   \\
\midrule
\multirow{6}{*}{\begin{tabular}[c]{@{}c@{}}\textbf{Token} \\\textbf{Add}\end{tabular}}& \textbf{Malicious URL} & 0 & 81\% & 0.7337 & 0 & 19\% & 0.6708 & \multirow{5}{*}{0.04/12.69} & 0.16/15.07 & 0 & \multirow{5}{*}{0.04/11.07} & 0.13/11.29  & 0  \\
& \textbf{Misleading prefix } & 0 & 83\% & 0.7051 & 0 & 33\% & 0.7432 &   & 0.17/15.56 & 0 &  & 0.93/11.06  & 0    \\
& \textbf{Misanthropic word} & 0 & 82\% & 0.7057 & 0 & 62\% & 0.6122 &  & 0.12/16.75  & 0 &   & 0.22/11.94 & 0  \\
& \textbf{Denied suffix} & 0 & 77\% & 0.7005 & 0 & 36\% & 0.5715 &  & 0.22/16.40  & 0 &  & 0.16/11.45 & 0  \\
& \textbf{Fictional content} &  0 & 86\% & 0.6842 & 0 & 85\% & 0.5445 &  & 0.69/21.13 & 0 &  & 0.10/11.90 & 0     \\ \cmidrule{2-14}
& \textbf{Average} &  0 & 81.8\% & 0.7058 & 0 & 47\% & 0.6284 & 0.04/12.69 & 0.27/16.98 & 0 & 0.04/11.07 & 0.31/11.53 & 0   \\
\bottomrule
\end{tabular}}
\end{table*}

%% file: tables/table4.tex
\begin{table}[]
\centering
\caption{Evaluations of target tokens with different similarities in token-substitution attack.}
\label{tab_target_token}
\resizebox{0.99\linewidth}{!}{\begin{tabular}{c|c|c|c|c|c}
\toprule
\textbf{Target} & \textbf{Similarity} & \textbf{BP} & \textbf{ASR-C} & \textbf{ASR} & \textbf{ATS} \\ \midrule
cat & 0.7609 & 5.63/31.56 & 1\% &  98\% & 0.7613 \\
wolf & 0.4482 & 5.16/30.14 & 2\% &  98\% & 0.7461 \\
elephant & 0.4060 & 5.96/30.94 & 0 & 99\%  & 0.7358  \\
bear & 0.3661 & 5.65/30.72 & 0 & 97\% & 0.7626  \\
tree & 0.2898 & 5.46/30.90 & 5\% &  99\% & 0.7568  \\
desk & 0.1228 & 5.81/30.97 & 1\% & 97\% & 0.7283  \\
\bottomrule
\end{tabular}}
\end{table}

%% file: tables/table8.tex
\begin{table}[t]
\centering
\caption{Impact of loss terms.}
\label{tab_loss}
\resizebox{0.95\linewidth}{!}{\begin{tabular}{c|c|c|c|c|c}
\toprule
\textbf{Attack} & \textbf{Removed} & \textbf{BP} & \textbf{ASR-C} & \textbf{ASR} & \textbf{ATS} \\ \midrule
\multirow{4}{*}{Token Sub}& $L_{bd}$ &  6.75/31.00 &  2\% & 2\% & 0.9150 \\
& $L_{cl}$ & 4.82/29.48  & 90\% & 100\% & 0.8722  \\
& $L_{emb}$ & 1.37/25.04 &  2\% & 97\% & 0.5958  \\
& None &  5.63/31.56 & 1\% & 98\% & 0.7613  \\
\midrule
\multirow{4}{*}{Token Add} & $L_{bd}$ &  3.94/29.69 & 0 & 0 & 0.8841 \\
& $L_{cl}$ & 3.71/27.64  &  100\% & 100\% & 0.8285  \\
& $L_{emb}$ & 3.12/28.83 & 1\% & 100\% & 0.8294  \\
& None &  3.41/29.29 & 0 & 100\%  & 0.8234 \\
\bottomrule
\end{tabular}}
\end{table}

%% file: tables/table12.tex
\begin{table}[t]
\renewcommand{\arraystretch}{0.5}
\centering
\caption{Different instruction templates for evaluation.}
\label{tab_prompt_detail}
\begin{tabular}{c|p{5.2cm}}
\toprule
\textbf{Type}                 & \textbf{Prompt}         \\ \midrule
Template 1           & $\langle$\texttt{image}$\rangle$$\backslash$n Describe the image in detail.    \\ \midrule
Template 2             & $\langle$\texttt{image}$\rangle$$\backslash$n Generate a descriptive caption for the image provided. \\ \midrule
Template 3    & $\langle$\texttt{image}$\rangle$$\backslash$n Create an engaging and imaginative caption for the given image.                                                                     \\ \midrule
Template 4              &  $\langle$\texttt{image}$\rangle$$\backslash$n  Craft an emotionally resonant caption for the provided image.    \\              
\bottomrule

\end{tabular}
\label{tab_prompt_detail}
\end{table}

%% file: tables/table7.tex
\begin{table}[t]
\centering
\caption{Impact of shadow dataset size.}
\label{tab_shadow_size}
\resizebox{0.95\linewidth}{!}{\begin{tabular}{c|c|c|c|c|c}
\toprule
\textbf{Attack} & \textbf{Size} & \textbf{BP} & \textbf{ASR-C} & \textbf{ASR} & \textbf{ATS} \\ \midrule
\multirow{3}{*}{Token Sub}& 500 &  5.22/30.19 &  35\% & 45\% & 0.8521 \\
& 1000 & 5.13/30.21  &  3\% & 97\% & 0.7489  \\
& 2000 &  5.63/31.56 & 1\%  & 98\% & 0.7613  \\
\midrule
\multirow{3}{*}{Token Add} & 500 &  2.15/22.27 &  65\% & 68\% & 0.5805 \\
& 1000 & 4.07/29.39  &  0 & 96\% & 0.7958 \\
& 2000 &  3.41/29.29 & 0 & 100\% & 0.8234  \\
\bottomrule
\end{tabular}}
\end{table}

%% file: tables/table16.tex
\begin{table}[t]
\caption{ASRs on two real world scenes.}
\label{tab_real_world}
\resizebox{0.95\linewidth}{!}{\begin{tabular}{ccc|cc}
\toprule
\multirow{2}{*}{Attack Scenes} & \multicolumn{2}{c|}{Clean Model} & \multicolumn{2}{c}{Backdoor Model} \\ \cmidrule{2-5} 
 & Clean & Triggered & Clean & Triggered \\ \midrule
Autonomous Driving & 0 & 0 & 0 & 100\% \\ \midrule
Medical Diagnosis & 0 & 0 & 2\% & 98\% \\ \bottomrule
\end{tabular}}
\end{table}

%% file: tables/table14.tex
% \begin{table}[t]
% \centering
% \caption{\bf Fine-tuning-based defense with different clean dataset size against BadToken.}
% \label{tab_defense_finetuning_size}
% \resizebox{0.95\linewidth}{!}{\begin{tabular}{c|c|c|c|c|c}
% \toprule
% \textbf{Attack} & \textbf{Size} & \textbf{BP} & \textbf{ASR-C} & \textbf{ASR} & \textbf{ATS} \\ \midrule
% \multirow{3}{*}{Token Sub}& 500 & 5.42/31.39 &  8\% & 98\% & 0.7697 \\
% & 1000 &  6.16/31.78 &  15\% & 96\% & 0.7696  \\
% & 2000 & 6.61/32.28  & 30\%  & 87\% & 0.7763  \\
% \midrule
% \multirow{3}{*}{Token Add} & 500 &  2.15/22.27 &  0 & 0 & 0.5805 \\
% & 1000 & 4.07/29.39  &  0 & 0 & 0.7958 \\
% & 2000 &  3.41/29.29 & 0 & 0 & 0.8234  \\
% \bottomrule
% \end{tabular}}
% \end{table}

\begin{table}[t]
\centering
\caption{Fine-tuning-based defense with different clean dataset size against BadToken.}
\label{tab_defense_finetuning_size}
\resizebox{1.0\linewidth}{!}{\begin{tabular}{c|c|c|c|c|c|c|c|c}
\toprule
\multirow{2}{*}{\textbf{Size}} & \multicolumn{4}{c}{\textbf{\textbf{Token Sub}}} & \multicolumn{4}{c}{\textbf{\textbf{Token Add}}} \\
\cmidrule(lr){2-5} \cmidrule(lr){6-9}
& \textbf{BP} & \textbf{ASR-C} & \textbf{ASR} & \textbf{ATS} & \textbf{BP} & \textbf{ASR-C} & \textbf{ASR} & \textbf{ATS} \\ \midrule
\textbf{500} &  5.42/31.39 &  8\% & 98\% & 0.7697 & 2.15/22.27 &  0 & 0 & 0.5805 \\
\textbf{1000} & 6.16/31.78 &  15\% & 96\% & 0.7696 & 4.07/29.39  &  0 & 0 & 0.7958 \\
\textbf{2000} & 6.61/32.28  & 30\%  & 87\% & 0.7763 & 3.41/29.29 & 0 & 0 & 0.8234 \\
\bottomrule
\end{tabular}}
\end{table}